%% file: paper.tex
\documentclass[letterpaper,twocolumn,10pt]{article}
\usepackage{usenix2019_v3}

\usepackage{cite}
\usepackage{url}
\usepackage{filecontents}
\usepackage{color}
\usepackage{multirow}
\usepackage{url}
\usepackage{graphicx}
\usepackage{algorithm}
\usepackage[noend]{algpseudocode}
\usepackage{subfigure}
\usepackage{booktabs}

\usepackage{enumitem} 
\usepackage{times}
\usepackage{xspace}
\usepackage{latexsym}
\usepackage{amsmath}
\usepackage{amsfonts}
\usepackage{tikz}
\usepackage{tabularx}
\usepackage{balance}
\usepackage{pifont}
\usepackage{soul}
\usepackage{siunitx}
\usepackage{amsthm}
\usepackage{bbm}

\setitemize{noitemsep,topsep=1pt,parsep=1pt,partopsep=1pt}
\setenumerate{noitemsep,topsep=1pt,parsep=1pt,partopsep=1pt}

\newcommand{\parabf}[1]{\smallskip\noindent\textbf{#1}}
\newcommand{\parait}[1]{\medskip\noindent\textit{#1}}
\newcommand{\paraf}[1]{\noindent\textbf{#1}}
\newcommand{\cut}[1]{}

\newcommand{\sysname}{SeaLLM\xspace}

\newcommand{\revise}[1]{#1}

\usepackage{array}
\newcommand{\PreserveBackslash}[1]{\let\temp=\\#1\let\\=\temp}
\newcolumntype{C}[1]{>{\PreserveBackslash\centering}p{#1}}
\newcolumntype{R}[1]{>{\PreserveBackslash\raggedleft}p{#1}}
\newcolumntype{L}[1]{>{\PreserveBackslash\raggedright}p{#1}}

\DeclareMathOperator*{\argmin}{arg\,min}

\newtheorem{theorem}{Theorem}

\newtheorem{lemma}{Lemma}

\begin{document}

\date{}

\title{\sysname: Service-Aware and Latency-Optimized Resource Sharing for \\
Large Language Model Inference}


\author{
{\rm Yihao Zhao}\\
Peking University
\and
{\rm Jiadun Chen}\\
Huawei Cloud
\and
{\rm Peng Sun}\\
Shanghai AI Lab
\and
{\rm Lei Li}\\
Peking University
\and
{\rm Xuanzhe Liu}\\
Peking University
\and
{\rm Xin Jin}\\
Peking University
}

\maketitle

\input{sections/abstract}
\input{sections/introduction}
\input{sections/motivation}
\input{sections/architecture}
\input{sections/design}
\input{sections/implementation}
\input{sections/evaluation}
\input{sections/related}
\input{sections/conclusion}

\label{lastpage}

\bibliographystyle{plain}
\bibliography{paper}

\input{sections/appendix}
\label{wholepage}

\end{document}

%% file: sections/abstract.tex
\begin{abstract}
Large language models (LLMs) with different architectures and sizes have been developed.
\revise{
Serving each LLM with dedicated GPUs leads to resource waste and service inefficiency due to the varying demand of LLM requests.
A common practice is to share multiple LLMs.
}
However, existing sharing systems either do not consider the autoregressive pattern of LLM services, or only focus on improving the throughput, which impairs the sharing performance, especially the serving latency.
We present \sysname, which enables service-aware and latency-optimized LLM sharing.
\sysname improves the overall sharing performance by
(1) a latency-optimized scheduling algorithm utilizing the characteristics of LLM services,
(2) a placement algorithm to determine the placement plan and an adaptive replacement algorithm to decide the replacement interval,
and (3) a unified key-value cache to share GPU memory among LLM services efficiently.
Our evaluation under real-world traces and LLM services demonstrates that \sysname improves the normalized latency by up to $13.60\times$, the tail latency by up to $18.69\times$, and the SLO attainment by up to $3.64\times$ compared to existing solutions.
\end{abstract}

%% file: sections/introduction.tex
\section{Introduction}
\label{sec:intro}

The fast development of large language models (LLMs) is changing modern applications, including chatbot~\cite{chiang2023vicuna}, code generation~\cite{li2022competition}, 
text process~\cite{brown2020language}, and embodied AI~\cite{song2023llm}. Many organizations~\cite{dubey2024llama, brown2020language, achiam2023gpt,zhang2022opt}
have proposed their foundation LLMs with different architectures and model sizes.
Service providers usually deploy different LLMs for specific tasks~\cite{duan2024muxserve,wu2024dlora,intel2024iks}.
However, LLM services have a higher demand for GPU resources in computation, memory, and communication than classical deep learning (DL) models.
The service providers need clusters with high-performance GPUs to provide high-quality LLM services, leading to high costs in building, using, and maintaining the clusters. 

\begin{figure}[t]
    \centerline{\includegraphics[width=\linewidth]{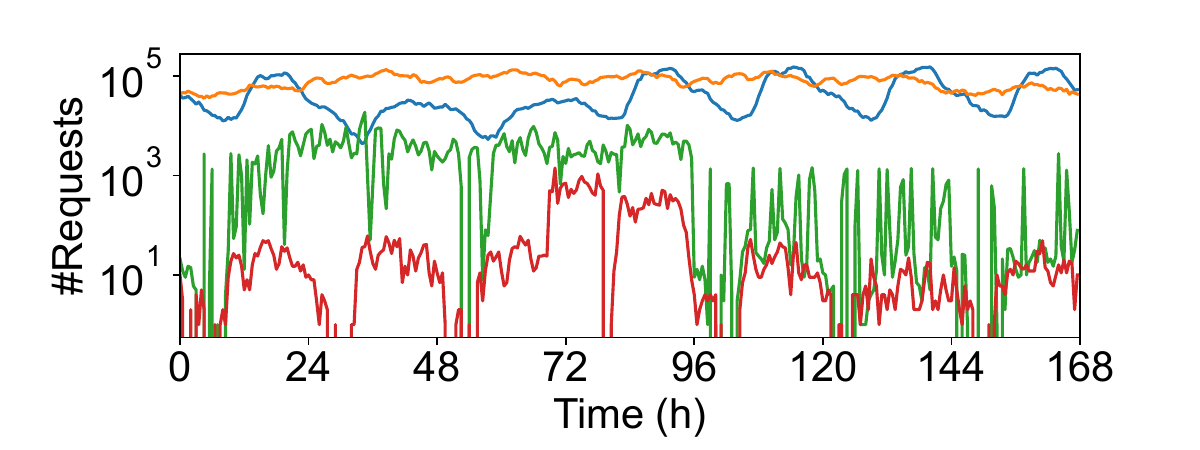}}
    \vspace{-0.15in}
    \caption{The request traffic of four production LLM services during seven days~\cite{stojkovic2024dynamollm, wang2024towards}.}
    \vspace{-0.1in}
    \label{fig:motiv_traffic}
\end{figure}

The request traffic of LLM services is dynamic for different time and services according to the data from production LLM services~\cite{stojkovic2024dynamollm, wang2024towards} as shown in Figure~\ref{fig:motiv_traffic}. 
Therefore, serving each LLM with dedicated GPUs leads to substantial wastes of expensive GPU resources and impacts the serving performance~\cite{li2023alpaserve,duan2024muxserve}.
Some systems~\cite{han2022microsecond, NVIDIA2022,li2023alpaserve,duan2024muxserve} share resources to improve serving performance and resource utilization.

Most existing sharing systems~\cite{han2022microsecond, NVIDIA2022,li2023alpaserve} are designed for serving classical DL models like the convolutional neural network (CNN) and they usually use the first-come-first-serve (FCFS) scheduling algorithm.
However, LLM services have a unique autoregressive pattern, i.e., generating the output tokens one by one.
The autoregressive pattern results in various output lengths and serving time for different LLM services and requests.
Therefore, the FCFS scheduling algorithm faces the head-of-line blocking problem for the shared LLM services, where a long request can block the following requests and affect the overall serving performance~\cite{kaffes2019shinjuku,wu2023fast}.

Recently, MuxServe~\cite{duan2024muxserve} is proposed for sharing LLM services.
MuxServe utilizes spatial and temporal multiplexing to improve the system throughput.
However, it influences the request latency, which is important to LLM services~\cite{zhong2024distserve}.
The reasons are mainly two-fold.
First, MuxServe adopts the round-robin (RR) scheduling algorithm for the prefill phase and the FCFS scheduling algorithm for each service.
The RR scheduling algorithm can avoid the head-of-line blocking problem~\cite{chen2019rrs,alizadeh2013pfabric}, but is still not adequate to serve multiple LLM services.
For long and bursty requests, the RR scheduling algorithm introduces a heavy burden on memory because LLM serving systems usually use the key-value cache (KV cache) to speed up serving and it requires storing the key-value pairs (KV pairs) of all requests.
Besides, the RR scheduling algorithm increases the latency for bursty requests.
Second, MuxServe predefines a quota of computation resources for each LLM service to enable spatial multiplexing.
The predefined quota cannot react to the dynamic request traffic timely, leading to GPU waste and an increase in latency.

This paper presents \sysname to support service-aware and latency-optimized resource sharing for LLM services.
The key observation is that LLM services usually have their specific characteristics, including the iteration time and the distributions of input and output lengths. 
Based on the characteristics, we place the LLM services and schedule the requests to alleviate the head-of-line blocking problem caused by FCFS and the burden of GPU memory caused by RR.

\sysname exploits a service-aware and latency-optimized scheduling algorithm for the shared LLM services.
We assign a priority based on the executed time and the service characteristics to each request.
Then we schedule the requests preemptively according to the priority.
The scheduling algorithm is proved to minimize the normalized latency with accurate profiling information.
However, in practice, some requests may deviate from the service characteristics and break the optimality of the scheduling algorithm.
Intuitively, for the overall performance, if one request exceeds the estimated execution time, its priority should be decreased to allow other requests to be processed.
With this intuition, we propose a doubling budget (DB) scheduling algorithm, which assigns a budget of execution time based on the profiled characteristics of each LLM service.
The DB scheduling algorithm gradually decreases the priority of one request when its budget is run out and doubles the budget to continue serving the request.

\revise{
\sysname leverages a search-based placement algorithm to generate the placement plan, i.e., how to place the LLM services in a cluster.
The time complexity of finding the optimal placement plan grows exponentially along with the cluster size and the number of LLM services.
}
To address this problem, we separate the search process into two stages, i.e., GPU group partition and LLM service allocation.
We also propose several heuristics to reduce the search time further.
With parallel execution, we can decrease the search time to minutes and fully hide the search process with service execution.
Besides, due to the time-varying request traffic, a fixed replacement is not a panacea.
Commonly, the solution to this problem is to replace the placement periodically~\cite{xiao2020antman,zhao2022multi}.
However, prior methods leverage a fixed replacement interval, which is hard to decide.
\sysname adopts an adaptive replacement algorithm, which changes the replacement interval according to the estimated performance and the real-achieved performance.

We implement \sysname as a cluster system with both a cluster manager to orchestrate the LLM services and requests, and local engines to execute the shared LLM services.
For the shared LLMs, memory management is a knotty problem, especially the dynamic KV cache.
A cluster system should share the KV cache of different LLM services to improve the serving performance~\cite{duan2024muxserve}.
However, the cache block shape is determined by the LLM architecture and is not identical for all LLMs, making it difficult to share the KV cache.
\sysname utilizes a unified KV cache with merged cache blocks to balance the block table size, locality, and fragmentation.

We evaluate \sysname on a 32-GPU cluster with real-world traces~\cite{shahrad2020serverless} and widely-used LLMs~\cite{touvron2023llama2,zhang2022opt}.
The experiment results show that \sysname improves the normalized latency by up to $13.60\times$, the tail latency by up to $18.69\times$, and the service level objective (SLO) attainment by up to $3.64\times$ compared to existing state-of-the-art (SOTA) systems.
\revise{
At the token level, \sysname improves the average time-to-first-token (TTFT) by up to $51.98\times$ and reaches a similar average time-per-output-token (TPOT) with SOTA systems.
}

In summary, this paper makes the following contributions:
\begin{itemize}[leftmargin=*]
    \item We identify the limitations of existing sharing solutions and propose to utilize the characteristics of LLM services for service placement and request scheduling.
    \item We propose a latency-optimized scheduling algorithm utilizing the service characteristics for shared LLM services.
    \item We propose a placement algorithm and an adaptive replacement algorithm to determine the placement plan.
    \item We conduct a comprehensive evaluation of \sysname with real-world traces and LLM services.
\end{itemize}

%% file: sections/motivation.tex
\section{Background and Motivation}
\label{sec:motiv}

\subsection{LLM Service}
LLMs have been widely deployed for different tasks, such as chatbot, search engine, text process, etc~\cite{chiang2023vicuna,brown2020language}.
In practice, the foundation LLMs are usually finetuned with domain-specific data to serve specific tasks~\cite{wu2024dlora}. 
The LLM developed and deployed for a specific task is called an LLM service.

LLMs generate the output in an \textit{autoregressive pattern}, i.e., the LLM generates one token in one step and the process is repeated until meeting the termination conditions.
Naturally, the generation process is separated into two phases: prefill phase and decoding phase.
The prefill phase processes all the input tokens concurrently and gets the first output token.
The decoding phase only generates one output token in each step.
To avoid the redundant computation brought by the attention layers, LLMs usually adopt \textit{KV cache} \cite{pope2023efficiently} to cache the intermediate data, i.e., the KV pairs, in the GPU memory.
The KV cache is formed by multiple cache blocks and the block shape is decided by the number of attention layers, the number of attention heads, and the hidden sizes.
To conclude, LLM services have two main characteristics.
First, the serving latency varies with different model architectures, input lengths, and output lengths.
Second, the serving process requires large amounts of GPU memory to store the KV pairs.

LLM serving systems usually aim to reduce the serving latency, improve the SLO attainment, and improve the serving throughput.
To achieve these goals, existing LLM serving systems leverage techniques including kernel-level optimization~\cite{aminabadi2022deepspeed, dao2022flashattention, hong2024flashdecoding}, batching~\cite{yu2022orca, fang2021turbotransformers}, disaggregation~\cite{zhong2024distserve}, model parallelism~\cite{yu2022orca, pope2023efficiently}, etc.
These techniques improve the performance of a single LLM service and they are orthogonal to the cluster setting which contains multiple LLM services.

\begin{figure}[t]
    \centerline{\includegraphics[width=\linewidth]{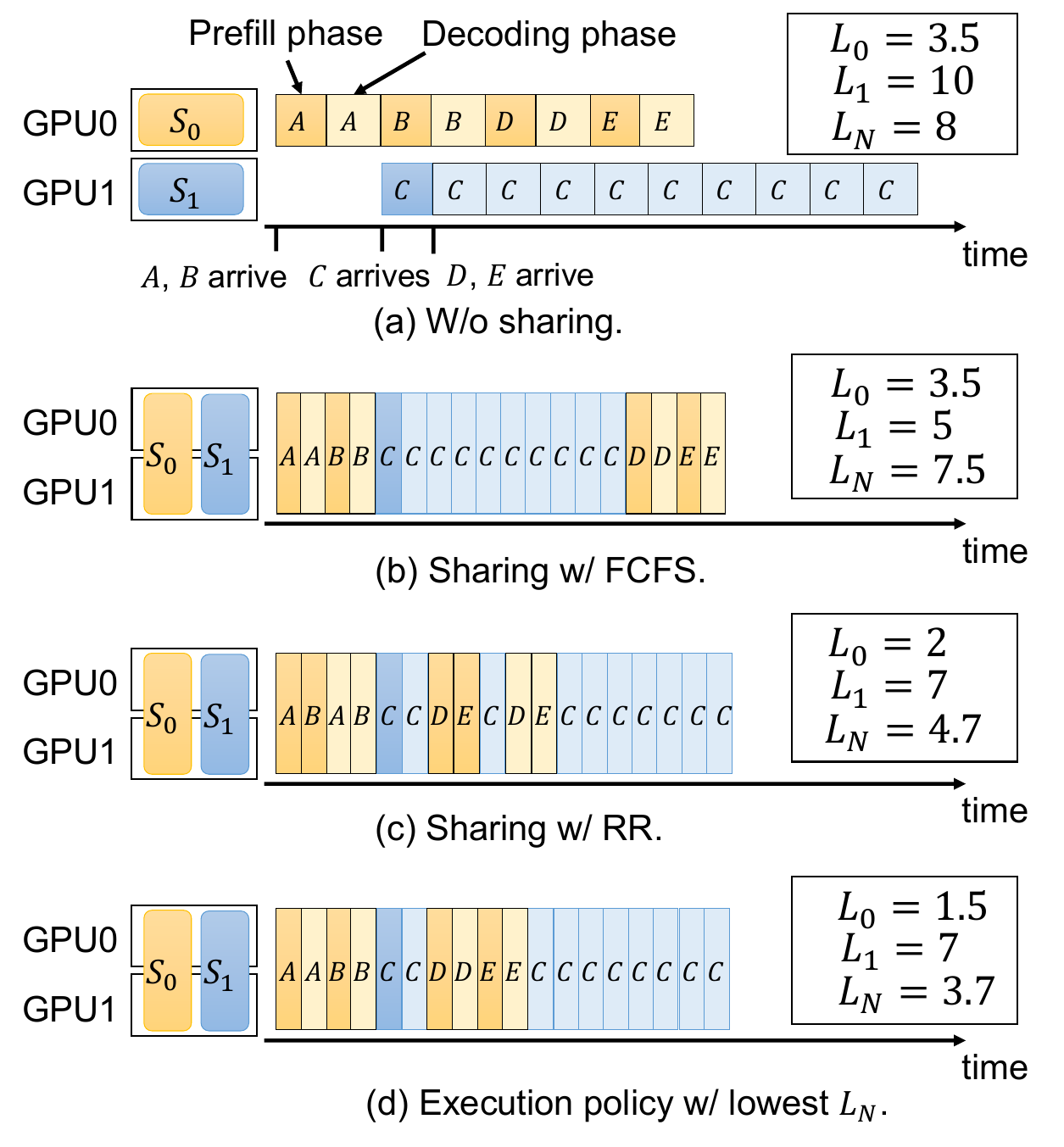}}
    \vspace{-0.15in}
    \caption{Four execution strategies of serving two LLM services on two GPUs.
    The blocks with dark colour are of the prefill phase and the blocks with light colour are of the decoding phase.
    $L_0$ and $L_1$ represent the average latency of service $S_0$ and $S_1$, respectively. $L_N$ represents the normalized latency of $S_0$ and $S_1$, defined in Equation~\ref{equ:norm_latency}.}
    \vspace{-0.1in}
    \label{fig:motiv_example}
\end{figure}

\subsection{LLM Sharing Systems}
Figure~\ref{fig:motiv_traffic} shows the dynamic nature of LLM services.
The data are from two public traces of four production LLM services~\cite{stojkovic2024dynamollm, wang2024towards}.
The request traffic of LLM services is time-varying and different services show different popularities.
The dynamic nature of LLM services brings opportunities for sharing LLM services to improve resource utilization and reduce costs for LLM service providers~\cite{li2023alpaserve,duan2024muxserve}.
Additionally, resource sharing can also benefit end users by improving the serving performance.
The example in Figure~\ref{fig:motiv_example} further illustrates the benefits of sharing LLM services.
For two LLM services ($S_0$ and $S_1$) and five requests ($A-E$), the execution strategy without sharing has many blanks, indicating the waste of GPU resources.
To better utilize the GPUs, we can collocate $S_0$ and $S_1$ on both GPUs by tensor parallelism as shown in Figure~\ref{fig:motiv_example}(b)-(d).
The average latencies of both services ($L_0$ for service $S_0$ and $L_1$ for service $S_1$) are decreased.

Most existing sharing systems~\cite{han2022microsecond, li2023alpaserve} are designed for classical models.
These models do not use the autoregressive pattern, i.e., for each request, the inference is only performed once.
Previous systems utilize the FCFS scheduling algorithm which has the lowest scheduling overhead.
However, as shown in Figure~\ref{fig:motiv_example}(b), the FCFS scheduling algorithm has the head-of-line blocking problem when applied to LLM services.
Specifically, the long request $C$ blocks the following requests $D$ and $E$, and influence the average latency of $S_0$.
As a result, the $L_0$ of FCFS is $2.3\times$ longer than the $L_0$ of (d).

Recently, MuxServe~\cite{duan2024muxserve} is proposed for sharing LLM services.
It utilizes the RR scheduling algorithm to select services for the prefill phase, and uses spatial multiplexing to share the decoding phase.
These mechanisms alleviate the problem of the FCFS scheduling algorithm.
However, it still influences the serving performance, especially the latency, because of two reasons.
First, the RR scheduling algorithm increases the request latency and the memory requirements for bursty requests.
As shown in Figure~\ref{fig:motiv_example}, (c) RR has a larger average latency $L_0$ than (d), because requests $D$ and $E$ are not finished as soon as possible.
Besides, (c) RR needs to store the KV pairs of at most three requests ($CDE$), while (d) only needs to store the KV pairs of at most two requests ($CD$ or $CE$).
The performance can be worse for more bursty requests.
Second, MuxServe assigns a quota of computation resources to each shared LLM service.
However, as the request traffic is dynamic, the preassigned quota increases the latency for bursty services and wastes GPUs for services at the traffic trough.
Although MuxServe improves the system throughput through sharing LLM services, it neglects the latency which is usually the primary goal of LLM serving systems.

\subsection{Challenges}
As we analyzed in \S 2.2, existing sharing methods do not perform well for LLM services.
We summarize the challenges in designing an efficient and latency-optimized sharing system for LLM services as follows.

\begin{table}[t]
\begin{small}
\centering
\begin{tabular}{ccrr}
\hline
Service         & Dataset                                 & \multicolumn{1}{c}{\begin{tabular}[c]{@{}c@{}}Avg input\\ length\end{tabular}} & \multicolumn{1}{c}{\begin{tabular}[c]{@{}c@{}}Avg output\\ length\end{tabular}} \\ \hline
Chatbot         & ShareGPT~\cite{ShareGPT2023}  & 73.0                                                                           & 426.9                                                                           \\
Summarization   & LongBench~\cite{bai2023longbench} & 13186.8                                                                        & 21.1                                                                            \\
Code & HumanEval~\cite{chen2021evaluating} & 156.5                                                                          & 66.9                                                                            \\ \hline
\end{tabular}
\vspace{-0.1in}
\caption{The average input length and output length of different LLM services.}
\vspace{-0.1in}
\label{table:motiv_length}
\end{small}
\end{table}

\begin{figure}[t]
	\begin{minipage}{0.48\linewidth}
        \centerline{\includegraphics[width=\linewidth]{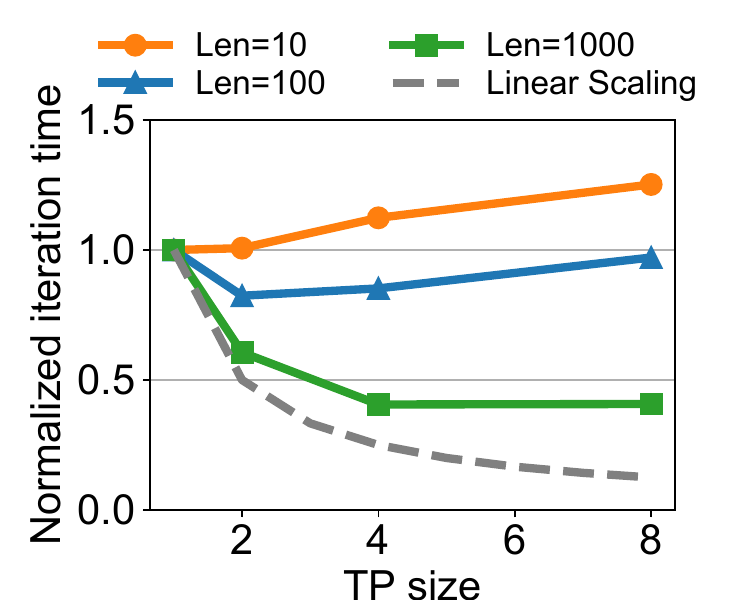}}
        \centerline{\small (a) Prefill phase, Llama2-7B.}
    \end{minipage}
    \begin{minipage}{0.48\linewidth}
        \centerline{\includegraphics[width=\linewidth]{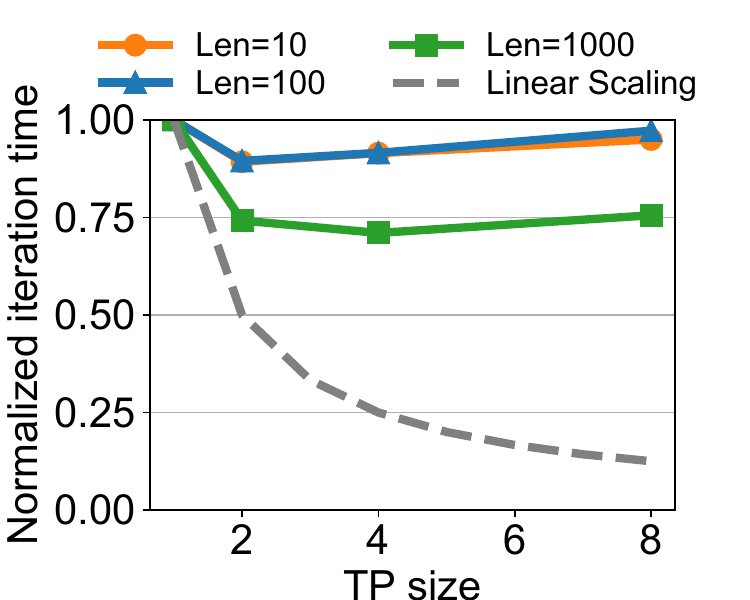}}
        \centerline{\small(b) Decoding phase, Llama2-7B.}
     \end{minipage}
     \begin{minipage}{0.48\linewidth}
        \centerline{\includegraphics[width=\linewidth]{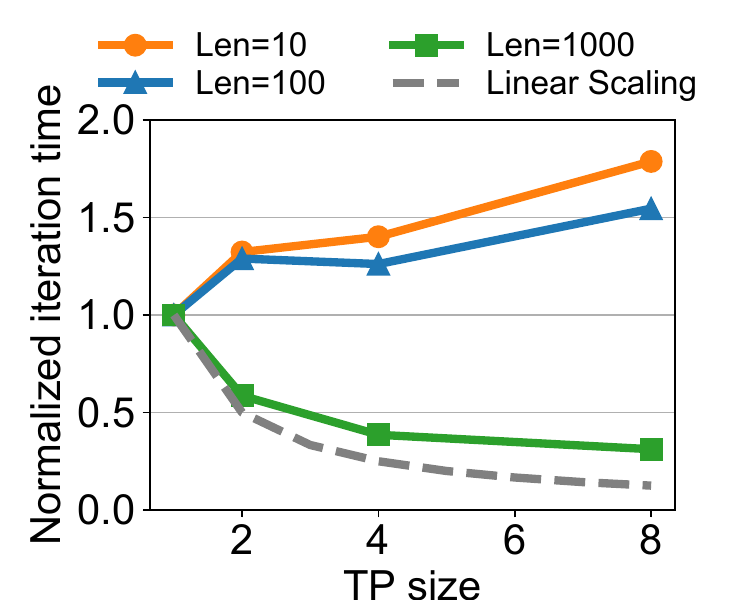}}
        \centerline{\small(c) Prefill phase, Llama2-13B.}
    \end{minipage}
    \begin{minipage}{0.48\linewidth}
        \centerline{\includegraphics[width=\linewidth]{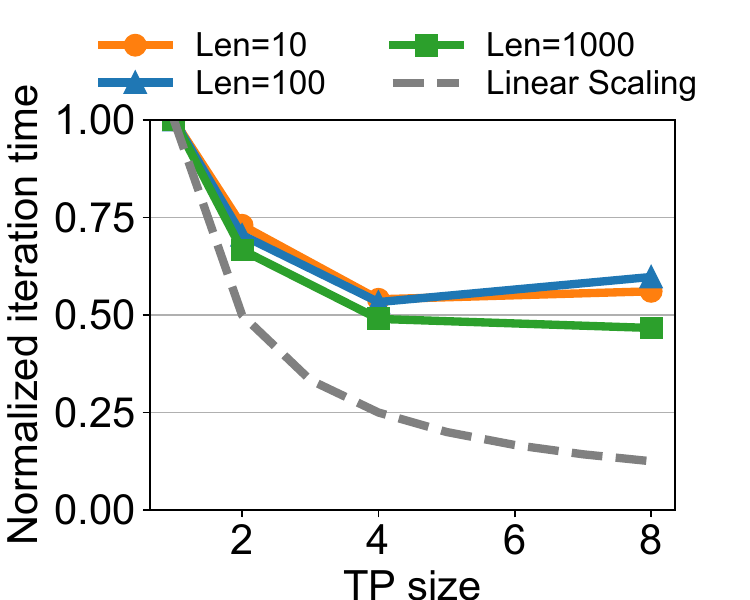}}
        \centerline{\small(d) Decoding phase, Llama2-13B.}
     \end{minipage}
    \caption{The iteration time of the prefill phase and the decoding phase under different TP sizes. 
    Len represents the number of input tokens and the batch size is 8.
    }
    \vspace{-0.1in}
    \label{fig:motiv_tp}
\end{figure}

\parabf{Various characteristics of LLM services.}
LLM services have various characteristics decided by the LLM architecture, and the distributions of the input and output lengths.
For example, Table~\ref{table:motiv_length} shows that the chatbot service (ShareGPT) has shorter inputs and longer outputs on average, while the summarization service (LongBench) has longer inputs and shorter outputs.
The optimal placement and parallelism plan can be different for different LLM services.
As shown in Figure~\ref{fig:motiv_tp}, a larger tensor parallelism (TP) size decreases the iteration time for long inputs served by Llama2-13B, while a larger TP size may increase the iteration time for short inputs served by Llama2-7B.
Besides, the speedups of TP are also different between the prefill phase and the decoding phase.
Hence, there is a critical need to consider the characteristics of LLM services for service placement and request scheduling.

\parabf{Large search space of service placement.}
For a cluster with multiple LLM services, allocating resources properly for each service is crucial for the overall cluster performance.
However, the search space of service placement is extremely large.
Even without sharing and parallelism, $n$ LLM services in a $m$-GPU cluster can have $m-1 \choose n-1$ kinds of possible placements, which is factorial to $n$ and $m$.
For example, there are more than $3\times 10^8$ placements for $16$ services in a $32$-GPU cluster.
Parallelism configuration influences the serving speed as shown in Figure~\ref{fig:motiv_tp}, and thus, we need to choose the proper parallelism configuration for each LLM service.
Additionally, LLM services are different in request traffic and the sharing group can affect the performance of each LLM service in it.
The parallelism configuration and the sharing group make the search space larger and more complex.
Therefore, how to place multiple LLM services efficiently is challenging.

\parabf{Compound requirements of scheduling.}
Efficiency and performance are two main requirements of the scheduling algorithm.
First, the scheduling algorithm should have low time complexity, i.e., efficiency.
We cannot design complex scheduling algorithms for LLM services as the inference process of requests only takes seconds or milliseconds.

Second, the scheduling algorithm should generate an execution sequence of requests with optimized serving performance.
Scheduling algorithms can influence the overall performance of all services in a cluster.
Note that the three scheduling algorithms in Figure~\ref{fig:motiv_example}(b)-(d) have different performances.
The main reason is that the characteristics of the services are different, where the requests of $S_0$ are more bursty and shorter than those of $S_1$.
Latency is one of the most important metrics for LLM services, because the output of LLM is long and users care about the response speed.
To better evaluate the overall performance, we compare the average latency of request $A-E$.
However, since the output lengths vary greatly, it is unfair to simply calculate the average latency of different requests.
Therefore, we use normalized latency, $L_N$, as a metric of overall performance, which can be formulated as,
\begin{equation}
\label{equ:norm_latency}
L_N = \sum_{s\in S}\sum_{r\in R_s} L_r/\hat{L}_s,
\end{equation}
where $S$ is the set of LLM services, $R_s$ is the requests of service $s$, $L_r$ is the real latency of request $r$, and $\hat{L}_s$ is the average request execution time of service $s$.
\revise{
The request latency $L_r$ can be decomposed into token-level metrics, including TTFT and TPOT.
Thus optimizing $L_N$ can also improve these token-level metrics to some extent.
}

As we analyzed in \S 2.2, the commonly used scheduling algorithms for serving workloads, i.e., FCFS and RR, are not suitable for shared LLM services.
As shown in Figure~\ref{fig:motiv_example}, the normalized latency $L_N$ of (d) is $2\times$ faster than that of FCFS and $1.3\times$ faster than that of RR.
Therefore, finding the latency-optimized execution order for the shared LLM services is both important and challenging.

\parabf{Large and complex memory requirements.}
LLMs require large GPU memory to store the KV cache, the model weights, and the intermediate data, which makes it difficult to serve multiple LLMs simultaneously on one GPU.
For example, a Llama2-7B model needs 14GB to store a half-precision model.
Besides, modern LLM serving systems usually leverage the KV cache, and the KV pairs for only one request with $4,096$ input tokens and $4,096$ output tokens need 4GB to store.
The large memory requirements of LLM services constrain the efficacy of resource sharing.
Moreover, the KV cache consists of multiple cache blocks and the block shape is decided by the LLM architecture.
As a result, we cannot simply share the KV cache of different LLM services.
We need to maintain the KV cache carefully for the shared LLM services.

%% file: sections/architecture.tex
\section{\sysname Overview}
\label{sec:arch}

\begin{figure}[t]
    \vspace{0.05in}
    \centerline{\includegraphics[width=\linewidth]{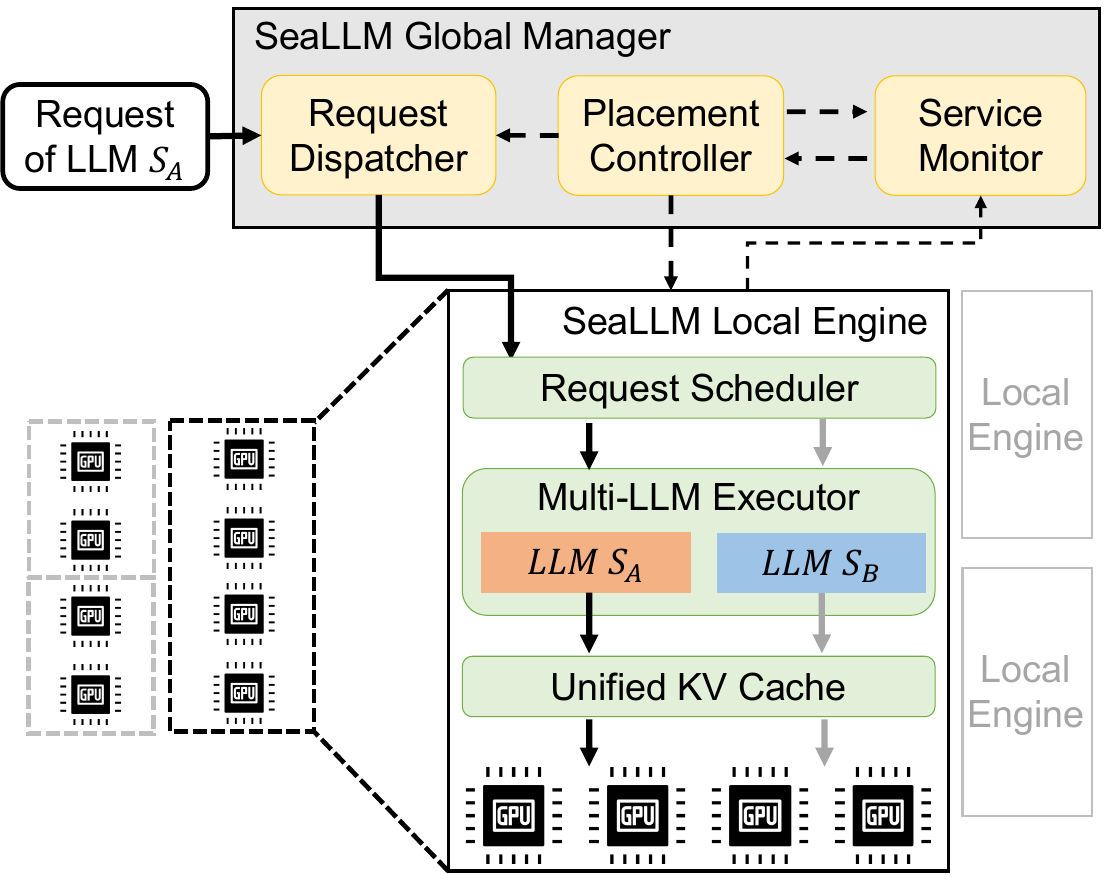}}
    \vspace{-0.15in}
    \caption{\sysname architecture.}
    \vspace{-0.1in}
    \label{fig:arch_arch}
\end{figure}

We propose an LLM service management system, \sysname, which enables efficient multi-LLM sharing, automatic LLM placement and replacement, and latency-optimized request scheduling.
Figure~\ref{fig:arch_arch} shows the overall architecture of \sysname.
\sysname consists of a global manager and a set of local engines on the GPU nodes.

\parabf{Global manager.}
\sysname's global manager is responsible for generating placement plans for LLM services, dispatching requests, and monitoring LLM services.
There are three components in the global manager, i.e., placement controller, request dispatcher, and service monitor.

\parait{Placement controller.}
The placement controller generates the placement plan for all the LLM services in the cluster.
A placement plan contains the used GPUs and the parallelism configuration for each LLM service.
As the request traffic of LLM services is time-varying, a fixed placement plan cannot react to the change of request traffic timely.
Therefore, the placement controller generates placement plans periodically based on the historical characteristics of services.
For a new LLM service, the placement controller allocates the minimum number of GPUs dedicatedly as an initial placement plan.
Note that a fixed replacement interval is not optimal for the system performance and is difficult to choose manually.
Specifically, a short replacement interval can lead to excessively frequent replacements, which increases system overhead.
In contrast, a long replacement interval cannot react to the dynamic request traffic timely, resulting in performance deterioration.
Hence, we propose an adaptive replacement algorithm to change the replacement interval automatically.

\parait{Service monitor.}
The service monitor monitors the LLM services in the cluster and collects their characteristics continuously.
The collected data include the model information (including the iteration time of prefill and decoding phases with different TP sizes), 
the request information (including the average and standard variance of the input and output lengths), 
and the execution information (including the latency of each request and the SLO attainment of each service).
These data are passed to the placement controller for periodical replacement and adjustment of the replacement interval.

\parait{Request dispatcher.}
When a new request arrives, the request dispatcher decides the proper local engine for the request according to its service type.
If multiple local engines can serve the arrived request, we select the local engine with the least number of requests.

\parabf{Local engine.}
\revise{
The local engine is responsible for scheduling and executing the requests dispatched by the global manager.
The local engines share the LLM services in time.
Each local engine consists of three components, i.e., request scheduler, multi-LLM executor, and unified KV cache.
}

\parait{Request scheduler.}
The request scheduler makes scheduling decisions for the dispatched requests.
There are two basic requirements for the scheduling algorithm.
\revise{
First, the scheduling algorithm should be fast, considering the short inference time of LLM requests.
}
Second, the policy should provide high serving performance.
Due to the unique characteristics of LLM services, simply applying FCFS or RR cannot reach optimal performance as shown in Figure~\ref{fig:motiv_example}.
We propose a DB scheduling algorithm by utilizing the service characteristics to achieve better scheduling performance.

\parait{Multi-LLM executor.}
Each batch of requests selected by the request scheduler is sent to the multi-LLM executor for execution.
The executor holds the shared models and invokes the corresponding model for inference.

\parait{Unified KV cache.}
In practice, the block shape of the KV cache is set according to the model architecture, e.g., the number of attention heads and layers, and the hidden size.
As a result, the block shape is not identical for LLMs.
\sysname exploits a unified KV cache to address this problem and the unified KV cache performs similarly to the KV cache for a single model.

%% file: sections/design.tex
\section{Design}
\label{sec:design}
The goal of \sysname is to provide latency-optimized resource sharing for multiple LLM services in a GPU cluster.
\revise{
Specifically, \sysname needs to decide how to place the LLM services in a given cluster and how to schedule the requests of each LLM service to improve the normalized latency.
}

First, we formulate the problem and our goal.
Assume there is a cluster with GPUs $G$ and multiple LLM services $S$.
For each service $s\in S$, its requests are represented as $R_s$ and its SLO is $SLO_s$.
We use \textit{sharing group} to represent a set of LLM services that share the same set of GPUs.
The placement of the $i$-th sharing group can be represented by a tuple $(S_i, G_i, p_i)$, where $S_i$ is the LLM services in this sharing group, $G_i$ is the GPUs used by this sharing group, and $p_i$ is the parallelism configuration for this group.
A placement plan $P$ for the cluster and all services $S$ can be defined as a set of placements for sharing groups, i.e., $P=\{(S_i, G_i, p_i)\}$.
With the above definition, our goal is to find the optimal placement plan $P^*$ and the execution order of requests $E^*$ to minimize the normalized latency under SLO constraints.
We formulate the problem as follows,
\begin{equation}
\label{equ:placement_problem}
P^*, E^*= \argmin_{P, E} L_N(S,P,E),
\end{equation}
where $L_N(\cdot,\cdot,\cdot)$ represents the normalized latency for services $S$ using placement plan $P$ and execution order $E$.
$L_N$ is defined in Equation~\ref{equ:norm_latency}.
Besides, we have three constraints,
\begin{equation}
\label{equ:all_services}
\cup S_i = S ,
\end{equation}
\begin{equation}
\label{equ:all_gpus}
\cup G_i \subseteq G ,
\end{equation}
\revise{
\begin{equation}
\label{equ:slo_constraint}
\sum_{r\in R_s} \mathbb{I}(L_r<SLO_s)/|R_s| \ge \Delta, \forall s\in S,
\end{equation}
where $\mathbb{I}$ is an indicator function and it is one when the following condition is satisfied.
}
Equation~\ref{equ:all_services} constraints that all LLM services should be placed in the cluster.
Equation~\ref{equ:all_gpus} constraints that the number of used GPUs should not exceed the number of all GPUs in the cluster.
Equation~\ref{equ:slo_constraint} denotes that the SLO attainment of all LLM services in the cluster should be larger than a threshold $\Delta$.

    
The number of all placement and execution orders is factorial to the number of GPUs and services.
Finding the optimal placement plan $P^*$ and execution order $E^*$ is a complex combinatorial optimization problem and it is almost impossible to find within an acceptable time.
Hence, we split the above problem into two stages, i.e.,
(1) generate the placement plan for each LLM service (\S4.1),
and (2) given the placement plan, decide the execution order of the requests submitted to each LLM service (\S4.2).
Additionally, we propose an adaptive replacement algorithm for the dynamic traffic (\S 4.3).
In the last, we introduce how we manage the unified KV cache for efficient service sharing (\S4.4).

\subsection{Service Placement}
A placement plan includes the sharing groups, the used GPUs for each sharing group, and the parallel configurations for each sharing group.
The number of all possible placements is factorial to the number of GPUs and services.
For example, a cluster with $32$ GPUs and $16$ services has more than $10^{8}$ possible placements even without considering the sharing groups and the parallelism configurations.
Thus, it is impossible to search all placements and find the optimal one within a reasonable time.
\revise{
We propose a two-stage placement algorithm similar to AlpaServe~\cite{li2023alpaserve}, but we extend it for LLM services.
First, we propose two heuristics to improve the searching speed because the cluster size and the number of parallel configurations of LLMs are getting larger than previous models.
Second, we utilize the normalized latency and our proposed latency-optimized scheduling algorithm to evaluate the searched placement plans.
}
Algorithm~\ref{alg:placement} in Appendix~\ref{sec:appendix_placement} is the pseudocode of our placement algorithm.
First, we partition the whole cluster into GPU groups by enumerating the parallelism configuration.
Then, we allocate LLM services to the GPU groups to form sharing groups.
With a simulator estimating the serving performance based on the historical request information, we can select the optimal placement with the lowest normalized latency.
The simulator uses our scheduling algorithm introduced in \S 4.2.

\parabf{Group partition.}
For the first stage, we enumerate all possible parallelism configurations to split the whole cluster.
Existing practice mainly uses tensor parallelism for LLM inference services~\cite{kwon2023efficient,wu2024dlora}, and thus we only consider tensor parallelism here.
\revise{
Note that our placement algorithm can be generalized to other parallelisms by enumerating their possible sizes.
}
Assume the cluster has $N$ nodes and each node has $m$ GPUs, i.e., the cluster has $Nm$ GPUs in total.
The partition space has $2^{Nm-1}$ partitions, which is almost impossible to enumerate for a large cluster.
We introduce two heuristics to prune the search space.
First, we only select the power of two as the parallelism size which is commonly used in practice~\cite{li2023alpaserve,zhong2024distserve}.
Besides, the tensor parallelism size cannot exceed $m$ because tensor parallelism needs large amounts of communication among GPUs and inter-node communication is slower than intra-node communication.
\revise{
\sysname can be generalized to cross-node serving by enabling other parallelisms, like pipeline parallelism.
}
Second, we set all groups with the same tensor parallelism size.
If the tensor parallelism size is too small for some services, we will merge two small groups into a larger one until all services can be placed.

\parabf{Service allocation.}
In the second stage, we allocate LLM services to the GPU groups partitioned by the first stage.
Similar to the first stage, enumerating all possible allocations is time-consuming.
Therefore, we use a heuristic to speed up the process.
We allocate services to the GPU group iteratively, and we select the most unserved service for each iteration.
We find that only using the normalized latency to choose the most unserved service is misleading and cannot satisfy the constraint of SLO attainment (Equation~\ref{equ:slo_constraint}).
Because during the allocation process, some services may not be allocated yet and their requests cannot be served, which leads to a fake small normalized latency.
To address this problem, we use the number of unserved requests as the main metric and the normalized latency as a secondary metric.
The unserved index $UI$ of service $s$ can be defined as follows,
\begin{equation}
\label{equ:unserved_index}
UI_s= \sum_{r\in R_s} \mathbb{I}(L_r>SLO_s)+\alpha L^s_N,
\end{equation}
where $L_N^s$ is the normalized latency of service $s$.
With the above definition of unserved index $UI_s$, \sysname allocates the service with the highest $UI_s$ to the GPU group which has the lowest request rate.
\revise{
We set $\alpha <<1$ to ensure that the SLO attainment is more important than the normalized latency.
This setting can improve the overall serving performance and avoid starvation of services with long output lengths.
}
When allocating an LLM service to a GPU group, \sysname first checks if the GPU memory is enough for serving the shared LLMs with profiled memory usage.

\subsection{Latency-Optimized Scheduling}
Given the placement, we need to decide the optimal execution order of requests for each LLM service, i.e., scheduling the requests of each LLM service to minimize the normalized latency.
There are two main requirements for the scheduling algorithm, i.e., efficiency and performance.
First, as the execution time of LLM services is only seconds or even milliseconds, the scheduling algorithm should be simple and fast, without bringing significant overhead for each request.
As a result, the complex scheduling algorithms for long-running tasks~\cite{zhao2022multi,GuZZXHCYHJL23} are not suitable for LLM services.

Second, the scheduling algorithm should provide high-performance scheduling decision for shared LLM services.
Existing LLM engines mainly use FCFS~\cite{kwon2023efficient,li2023alpaserve} or RR~\cite{duan2024muxserve,wu2023fast} schedul ing algorithms.
FCFS serves the requests in their arrival order.
It introduces almost no scheduling overhead for requests.
However, it is notorious for the head-of-line blocking problem, i.e., a long request in the head of the queue can block other requests.
This makes FCFS unsuitable for shared LLM services, especially when the services have highly diverse characteristics.
For example, in Figure~\ref{fig:motiv_example}, the requests of service $S_1$ have longer execution time than the requests of service $S_0$.
Using FCFS, the request $C$ blocks the requests $D$ and $E$, which greatly increases the latency of service $S_0$.
Some approaches use the RR scheduling algorithm.
They place the requests into different queues and execute them in a round-robin pattern.
These approaches can alleviate the head-of-line blocking problem, but can be inefficient for bursty and long requests with frequent preemptions.
\revise{
Although FastServe~\cite{wu2023fast} adopts the multi-level feed-back queue to avoid frequent preemptions, it does not consider the characteristics of each shared LLM service.
Thus, its improvement is limited when sharing LLM services.
}
Additionally, these approaches also impose a heavy burden on GPU memory, because many requests are executed in turn and the KV pairs of these requests should all be stored in GPU memory.

To address the above problems, we propose the doubling budget (DB) scheduling algorithm for shared LLM services.
Our key observation is that LLM services usually have different characteristics, e.g., the distributions of the input and output lengths, as shown in Table~\ref{table:motiv_length}.
Based on this observation, we can execute short requests first to decrease the normalized latency.
Specifically, we set the priority of request $r$ as 
\begin{equation}
\label{equ:priority}
O_r = T_r' \hat{L_r},
\end{equation}
where $T_r'$ is the estimated remaining time of $r$ and $\hat{L_r}$ is the profiled execution time of $r$ on one GPU.
Smaller $O_r$ represents higher priority and requests with higher priority are served first.
To avoid the head-of-line blocking problem, we use preemptive scheduling and schedule requests before each iteration. 
The proposed scheduling algorithm is proven to reach the optimal normalized latency under certain conditions.

\begin{theorem}
\label{thm:schedule}
\vspace{-0.05in}
The proposed scheduling algorithm can minimize the normalized latency with the following assumptions:
(1) requests can be preempted at any time, and (2) all requests of the same service have identical execution time.  
\vspace{-0.05in}
\end{theorem}

\begin{figure*}[t]
    \centerline{\includegraphics[width=0.9\linewidth]{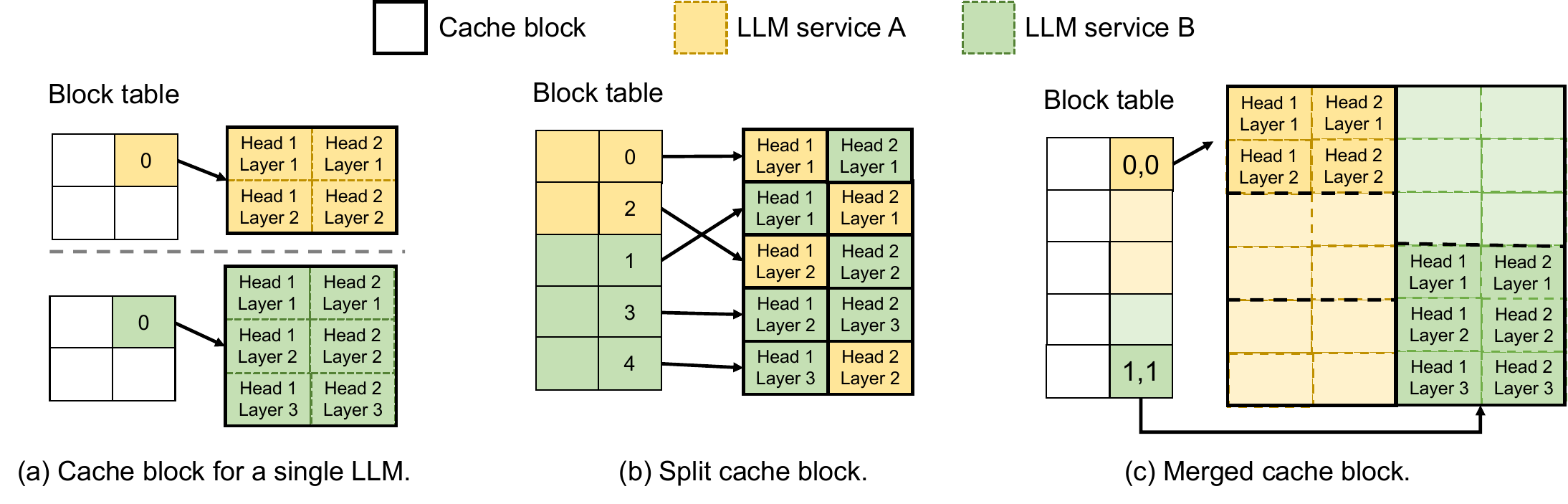}}
    \vspace{-0.2in}
    \caption{Comparison among (a) single LLM's cache block, (b) split cache block, and (c) merged cache block.}
    \vspace{-0.15in}
    \label{fig:design_kvcache}
\end{figure*}

The proof of Theorem~\ref{thm:schedule} is in Appendix~\ref{sec:appendix_schedule}. 
\revise{
Briefly, this problem is a special case of the queuing model $B|r_j, pmtn|\sum L_j/\hat{L_j}$, where $B$ is the batch size.
}
We can prove this theorem by contradiction.
Assumption (1) can be ignored as the execution time of one iteration is usually short, especially for the decoding phase.
\revise{
In practice, assumption (2) is usually not the case. 
However, our priority $O_r$ is still the optimal priority to minimize the expected normalized latency.
}

\begin{theorem}
\label{thm:schedule_prio}
\vspace{-0.05in}
For any scheduling moment, the proposed priority $O_r$ is optimal to minimize $\mathbb{E} (L_N)$.
\vspace{-0.05in}
\end{theorem}

\revise{
The proof of Theorem~\ref{thm:schedule_prio} is in Appendix~\ref{sec:appendix_schedule}.
However, in practice, some requests have extremely long outputs, leading to performance degradation.
}
For this, we modify our DB scheduling algorithm as shown in Algorithm~\ref{alg:db} in Appendix~\ref{sec:appendix_schedule}.
We introduce a budget of the execution time for each request and define the budget as $Q=(\hat{L_s}+Var_s)$, where $\hat{L_s}$ and $Var_s$ are the average and standard variance of the execution time of requests belonging to service $s$. 
The budget $Q$ decreases with the request execution.
The priority of a request $r$ is $O_r = Q_r\hat{L_s}$, where $r\in R_s$ and $Q_r$ is the left budget for request $r$.
We select the request with the highest priority, i.e., the smallest $O_r$, to execute first.
For better GPU utilization and throughput, we batch the requests in the same phase and of the same service.
If one request does not finish within its initial budget, we double its budget, i.e., $Q'=2(\hat{L_s}+Var_s)$, and reset its priority to $Q'\hat{L_s}$.
We repeat to double the budget once the budget is used up until the request is finished.
We set a starvation threshold for each service and serve the starved services first.
\revise{
Although our scheduling policy is probabilistically optimal, it may not reach the minimum normalized latency.
Besides, it is difficult to tell how far our scheduling decision is from the optimal scheduling decision, considering the unpredictable request lengths in practice.
}

\subsection{Adaptive Replacement}
The request traffic of LLM services is time-varying as shown in Figure~\ref{fig:motiv_traffic}.
A fixed placement plan cannot respond to the changing request traffic promptly, leading to performance deterioration when bursty requests come in.
To address this problem, we replace the LLM services in the cluster periodically.
Periodical replacement is not new for cluster management.
However, existing methods~\cite{zhao2022multi,xiao2020antman} usually adopt fixed replacement intervals, which are difficult to determine manually.
\revise{
Specifically, a short replacement interval increases the system overhead of migrating services, while a long replacement interval cannot provide new placement plans timely.
}

\sysname introduces an adaptive replacement algorithm to select the proper replacement interval automatically.
Our main idea is to change the replacement interval based on the difference between the achieved performance metrics and the estimated performance metrics.
\sysname has a replacement interval $I_t$ for the t-th replacement interval, where $I_0$ is an initial hyper-parameter selected by the cluster manager.
Specifically, before each interval, our simulator calculates the estimated performance metrics, including the SLO attainment and the normalized performance, based on the historical data of the last replacement interval.
During the execution, the service monitor records the performance metrics for each LLM service.
At the end of interval $t$, \sysname compares the real performance $M_{real}$ and the estimated performance $M_{est}$, and change the replacement interval $I_{t+1}$ as follows,
\begin{equation}
\label{equ:adaptive_replacement}
I_{t+1}=
\begin{cases}
(1-\beta)I_t& \text{$|M_{real}-M_{est}|/M_{est} >\beta$}\\
(1+\beta)I_t& \text{$|M_{real}-M_{est}|/M_{est} <\beta$},
\end{cases}
\end{equation}
where $\beta$ is a parameter in $(0,1)$.
Intuitively, a larger difference between $M_{real}$ and $M_{est}$ indicates the current request rates vary from historical ones, suggesting that we should shorten the replacement interval.
On the other hand, a smaller difference represents that the request rates remain relatively stable, allowing us to increase the replacement interval.

\subsection{Unified KV Cache}
Sharing multiple LLMs requires an efficient and flexible mechanism to manage the resources, especially the GPU memory.
GPU memory is mainly used to store three parts, i.e., the model weight, activation, and KV cache.
The model weight and the activation are relatively stable given the model information and the batch size.
We allocate sufficient space for both the model weights and the peak activation usage.
However, managing the KV cache for shared services is very challenging.
First, the KV cache is dynamic during model execution, and separating the KV cache for each LLM can cause memory fragmentation or memory scarcity.
Besides, the KV cache consists of multiple blocks whose shape is decided by the LLM architecture, including the number of hidden layers, the number of heads, and the hidden sizes.
Different LLM architectures result in different block shapes.
Therefore, we cannot directly share the KV cache of the shared services.

To efficiently manage the KV cache, \sysname introduces a unified KV cache mechanism, which changes the cache block shape to fit the shared LLMs.
The hidden size is usually identical for widely-used models~\cite{duan2024muxserve}, e.g., 128 for Llamas~\cite{touvron2023llama2} and OPT over 2.7B~\cite{zhang2022opt}.
For other dimensions, there are mainly two possible methods, i.e., split a larger block into smaller blocks (called split method) or merge smaller blocks into a larger block (called merged method).
Figure~\ref{fig:design_kvcache} shows an example of these two methods.
\sysname adopts the merged method mainly due to three reasons.
First, the merged method needs much less space for the block table than the split method.
For example, the split method needs an $1,024\times$ larger block table for Llama2-7B which has 32 hidden layers and 32 attention heads, and this number increases for larger LLMs.
On the other hand, the merged method only needs to add a second index for each original block to indicate the place in a merged block.
Second, the merged method requires fewer read/write operations to fetch/store the same amount of KV pairs.
Consequently, it achieves better locality in accessing the physical space compared to the split method, leading to improved read/write speeds.
Third, although the merged method may lead to more memory fragmentations than the split method, the batched execution can alleviate this fragmentation issue.

%% file: sections/implementation.tex
\section{Implementation}
\label{sec:impl}

\sysname is implemented with approximately 12,000 lines of code in Python, C++, and CUDA, and reuses some components of vLLM~\cite{kwon2023efficient}.
There are three roles in our system, i.e., the global manager, node worker, and local engine.

\parabf{Global manager.}
There is only one global manager for a cluster and the global manager is mainly implemented in Python.
It contains a component for receiving requests, the placement algorithm, the adaptive replacement algorithm, the request dispatcher, and the service monitor.
We run the placement algorithm in parallel as there are no dependencies for allocating models and evaluating the performance among different group patterns.
The global manager uses gRPC~\cite{GooglegRPC} to communicate with other roles in the cluster.
We use the Ray cluster for the distributed local engines.

\parabf{Node worker.}
There is one node worker for each node in the cluster and the node worker is mainly implemented in Python.
The node worker plays as a middleware to bridge the global manager and the local engine.
Besides, the node worker monitors the states of the local engines on the same node.
\revise{
When abnormal states happen, e.g., one service crashes and affects the shared services, the node worker will report to the global manager and help to restart the local engine.
}
The node worker collects service characteristics from local engines, and reports to the service monitor of the global manager.
When replacement happens, the node worker starts to pull model description files and weights from remote storage.
The pulling process is asynchronous to minimize the effect of running services.
\revise{
Once all nodes in the cluster are ready for the new placement plan, the node worker migrates ongoing requests after the ongoing iteration and stops old local engines.
}

\parabf{Local engine.}
Each group of shared LLM services has one local engine on the specific node.
The request scheduler is implemented in Python and uses the PriorityQueue whose complexities of adding a new element and getting the smallest element are both $O(\log n)$, where $n$ is the number of elements in the PriorityQueue.
The multi-LLM executor is based on the LLM engine of vLLM and is implemented in Python, C++, and CUDA.
We add support for sharing LLMs, including modifying the LLM engine and holding multiple model runners. 
The weights of the shared LLMs are all stored in the GPU memory for fast switch and execution.
For the unified KV cache, we implement a unified block manager in Python and C++.
The unified block manager can profile LLMs, calculate proper block shape, and manage the cache blocks for the shared LLM services.
We use Ray workers for parallel execution and use NCCL for communication.
The local engine also supports profiling and collecting the service characteristics, which are sent to the node worker and the global manager.

%% file: sections/evaluation.tex
\section{Evaluation}
\label{sec:eval}

\begin{table}[t]
\begin{small}
\centering
\begin{tabular}{cccc}
\hline
\textbf{Baseline}       & \textbf{Real testbed} & \textbf{Simulator} & \textbf{Difference} \\ \hline
vLLM                    & 13.03                  & 13.05                 & 0.2\%                   \\
AlpaServe                 & 11.20                  &   11.18             & 0.2\%                 \\
MuxServe                 & 19.55                  & 19.39               & 0.8\%                 \\
\sysname                & 1.65         & 1.60               & 3\%                 \\ \hline
\end{tabular}
\vspace{-0.1in}
\caption{Comparison of the normalized latency from real testbed and simulator.}
\label{table:simulator}
\end{small}
\vspace{-0.1in}
\end{table}

\begin{figure*}[t]
\begin{minipage}{0.32\linewidth}
    \centerline{\includegraphics[width=\linewidth]{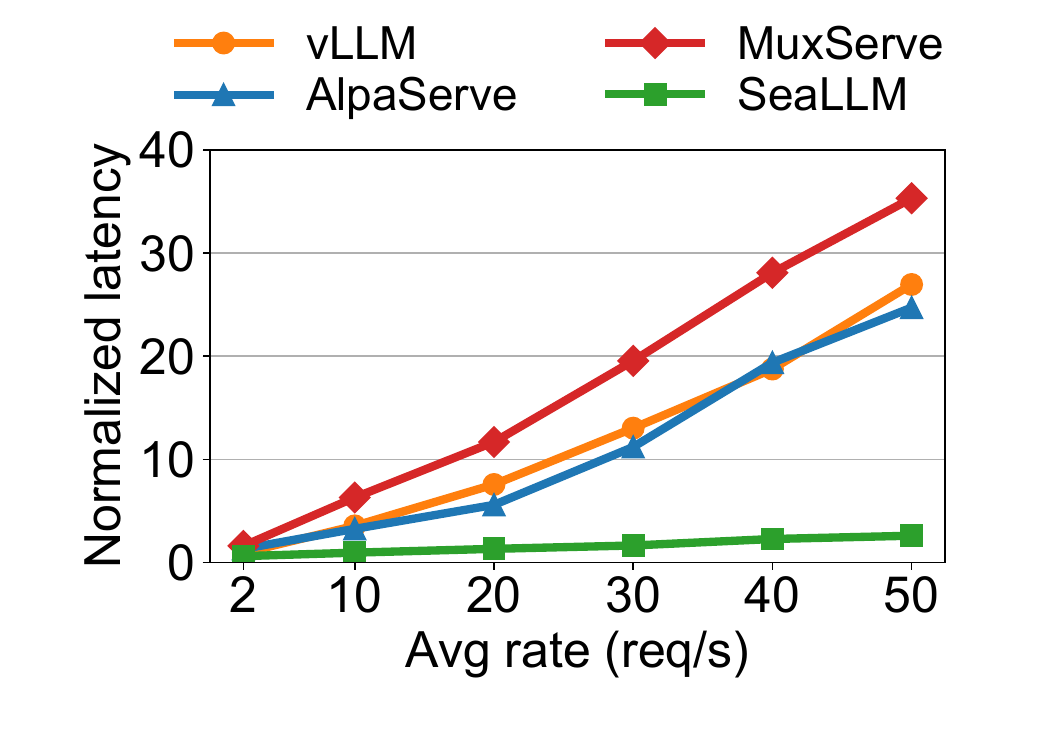}}
    \vspace{-0.15in}
    \centerline{\small (a) Normalized latency.}
    \vspace{-0.1in}
\end{minipage}
\begin{minipage}{0.32\linewidth}
    \centerline{\includegraphics[width=\linewidth]{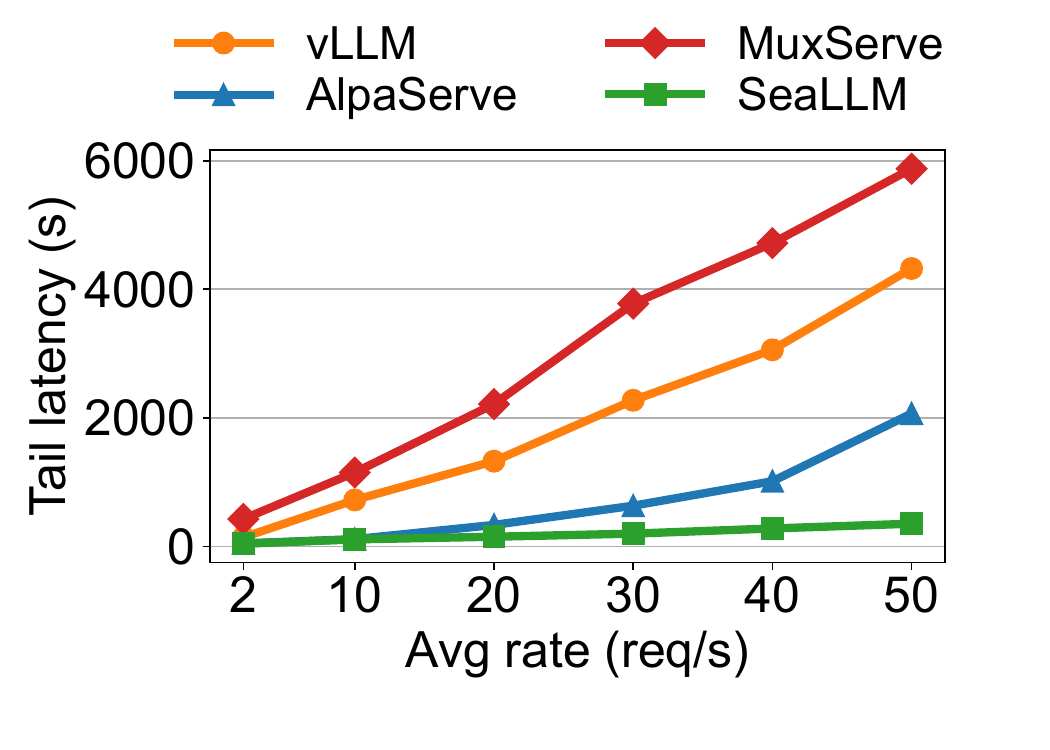}}
    \vspace{-0.15in}
    \centerline{\small (b) Tail latency.}
    \vspace{-0.1in}
\end{minipage}
\begin{minipage}{0.32\linewidth}
    \centerline{\includegraphics[width=\linewidth]{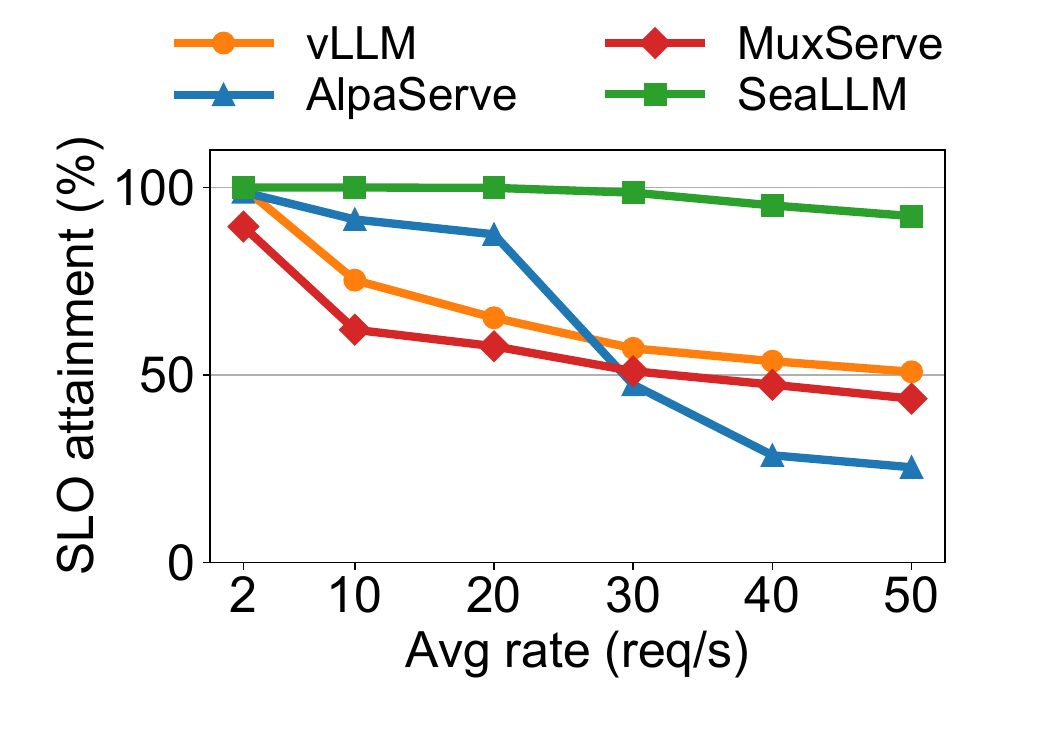}}
    \vspace{-0.15in}
    \centerline{\small (c) SLO attainment.}
    \vspace{-0.1in}
\end{minipage}
\caption{Testbed experiments on MAF traces.}
\label{fig:eval_testbed}
\vspace{-0.15in}
\end{figure*}

\begin{figure}[t]
\begin{minipage}{0.49\linewidth}
    \vspace{-0.1in}
    \centerline{\includegraphics[width=\linewidth,trim=0 0 0 30, clip]{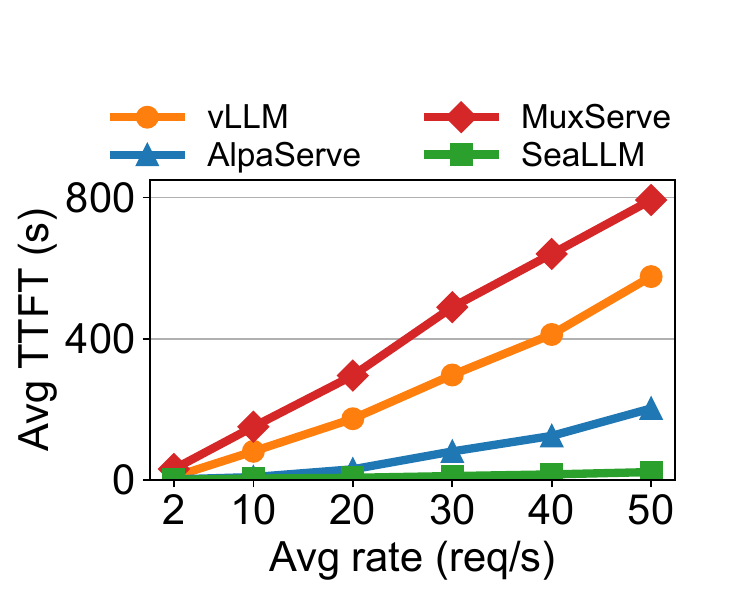}}
    \vspace{-0.05in}
    \centerline{\small (a) Avg TTFT.}
    \vspace{-0.05in}
\end{minipage}
\begin{minipage}{0.49\linewidth}
    \vspace{-0.1in}
    \centerline{\includegraphics[width=\linewidth,trim=0 0 0 30, clip]{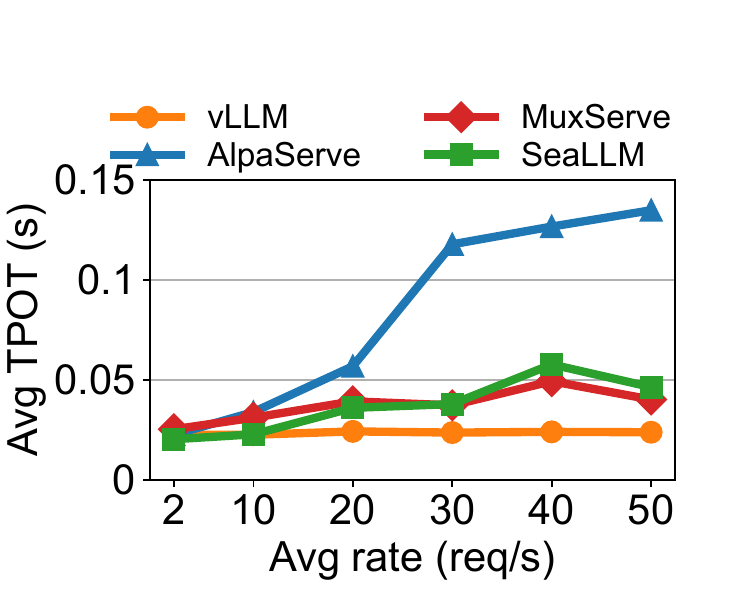}}
    \vspace{-0.05in}
    \centerline{\small (b) Avg TPOT.}
    \vspace{-0.05in}
\end{minipage}
\caption{Comparison of TTFT and TPOT.}
\label{fig:eval_ttft}
\vspace{-0.05in}
\end{figure}

We evaluate \sysname with real-world request traces and LLM services.
The evaluation shows that \sysname consistently outperforms the current SOTA systems.
Under different request rates, \sysname can improve the normalized latency by up to $13.60\times$, the tail latency by up to $18.69\times$, and the SLO attainment by up to $3.64\times$ (\S 6.2).
Moreover, we show the efficacy of the design choices of \sysname (\S 6.3) and we also analyze the overhead of \sysname (\S 6.4).

\subsection{Experiment Setup}

\paraf{Testbed.}
We conduct the testbed experiments on a cluster with 4 nodes and 32 GPUs.
Each node has 8 NVIDIA A800 80GB GPUs, 128 CPUs, 2048GB of host memory, four 200Gbps InfiniBand NICs, and 400GBps NVLink bandwidth between two GPUs.
We use PyTorch 2.1.2, CUDA 12.2, HuggingFace tokenizers 0.19.1, and Ray 2.35.0 for testbed experiments.
We set $\alpha$ in Equation~\ref{equ:unserved_index} to $0.0001$ and $\beta$ in Equation~\ref{equ:adaptive_replacement} to $0.1$.
\revise{
In our evaluation, we share at most two LLM services per local engine because of the GPU memory limitation.
}

\parabf{Simulator.}
Similar to prior work~\cite{zhao2022multi,li2023alpaserve}, we build a simulator to conduct the ablation study on a broader set of configurations.
We profile the prefill phase and the decoding phase for each LLM service and request under different TP sizes.
We also profile the memory usage of LLMs.
Table~\ref{table:simulator} shows that the differences between our simulator and real testbed are less than $3\%$, showing the high fidelity of our simulator.

\parabf{LLM service setup.}
Each LLM service is identified by the used LLM and the characteristics of requests.
We select four commonly used LLMs with model sizes ranging from 6.7B to 70B.
The selected LLMs are Llama2-7B\&13B\&70B~\cite{touvron2023llama2} and OPT-6.7B~\cite{zhang2022opt}.
The input lengths and output lengths of requests are sampled from the following real-world datasets:
(1) \emph{ShareGPT} dataset~\cite{ShareGPT2023} is a collection of conversations between users and ChatGPT. 
(2) \emph{LongBench} dataset~\cite{bai2023longbench} is a collection of summarization tasks.
Requests in the ShareGPT dataset have longer output lengths, while requests in the LongBench dataset have longer input lengths.
We use the K-means clustering algorithm to partition the requests according to their input and output lengths.
For each request group, we randomly assign an LLM to form the LLM service.

\parabf{Traces.}
Similar to prior work~\cite{li2023alpaserve,wu2024dlora}, we use the Microsoft Azure function trace (MAF)~\cite{shahrad2020serverless} for arrival patterns.
We round-robin functions in MAF to LLM services and generate traffic for each service.
We vary the request arrival rate and select requests from a fixed duration for each trace.

\parabf{Metrics.}
For each request rate, we measure the normalized latency (the average of end-to-end latency divided by the average execution time), the tail latency (the 99\%-th latency), and the SLO attainment (the percentage of requests that finish before their SLOs).
Lower normalized latency, lower tail latency, and higher SLO attainment represent better serving performance.
For the SLO attainment, we set the SLO to $5\times$ of the execution time for each request.
Besides, we also evaluate two widely used metrics for token-level serving performance, i.e., the average time-to-first-token (Avg TTFT) and the average time-per-output-token (Avg TPOT).

\parabf{Baselines.}
We compare \sysname with the following SOTA LLM serving systems:

\begin{itemize}[leftmargin=*]
    \item \emph{$vLLM$}~\cite{kwon2023efficient} is one of the most popular LLM serving systems for a single LLM. It cannot share services and uses FCFS for scheduling. We allocate the GPUs evenly to LLM services. 
    \item \emph{AlpaServe}~\cite{li2023alpaserve} proposes multiplexing for deep learning model inference. As it is not designed for LLMs with the autoregressive pattern, we implement AlpaServe with our unified KV cache and FCFS as the scheduling algorithm. 
    \item \emph{MuxServe}~\cite{duan2024muxserve} is designed for sharing LLM services. It uses RR and FCFS to schedule requests. It adopts the split method for the KV cache. We use its open-sourced code, which is also built upon vLLM, for evaluation.
\end{itemize}

\subsection{End-to-End Performance}
We first evaluate \sysname on a real cluster with $32$ GPUs.
Figure~\ref{fig:eval_testbed} shows the normalized latency, the tail latency, and the SLO attainment on the real-world MAF traces.
We change the average arrival rate of requests from 2 to 50 requests per second.
vLLM does not share LLM services and wastes GPU resources. 
AlpaServe uses the FCFS scheduling algorithm for the shared services, which has the head-of-line blocking problem and impairs the performance.
Compared to vLLM and AlpaServe, \sysname reaches up to $10.38\times$ and $9.52\times$ lower normalized latency, $12.13\times$ and $5.80\times$ lower tail latency, and $1.82\times$ and $3.64\times$ higher SLO attainment.

MuxServe shares LLMs in both time and space dimensions, but its performance is worse than \sysname, mainly due to three reasons.
First, MuxServe serves the services in the RR and FCFS manner, leading to lower performance when handling bursty requests.
Second, MuxServe sets a fixed percentage of computational resources for each LLM service, which means that it cannot adjust to fluctuating request traffic timely, resulting in wasted computational resources.
Third, MuxServe uses the split KV cache, which increases execution overhead.
Compared to MuxServe, \sysname utilizes the service characteristics for latency-optimized scheduling, adaptive replacement for proper placement plans, and merged KV cache for lower overhead.
\sysname improves the normalized latency by up to $13.60\times$, the tail latency by up to $18.69\times$, and the SLO attainment by up to $2.11\times$.

\revise{
The three metrics show that \sysname performs better than the SOTA baselines.
The normalized latency of \sysname is always lower than $3$ and the SLO attainment of \sysname is always higher than $90\%$ under different request rates, demonstrating the consistency of \sysname's improvement.
As the service characteristics are greatly different as shown in Table~\ref{table:motiv_length}, the lower tail latency and higher SLO attainment indicate that \sysname has better fairness for LLM services.
}

We also evaluate the Avg TTFT and the Avg TPOT, which can reflect the token-level serving performance, as shown in Figure~\ref{fig:eval_ttft}.
Avg TTFT is the time to get the first token and it is one of the most important metrics for online services.
\sysname can generate the first token up to $51.98\times$ faster than other baselines.
Avg TPOT is the average time of generating each output token.
Sharing LLMs inevitably increases the TPOT.
\sysname is always faster than the temporal sharing method, AlpaServe, showing \sysname's efficiency.
The TPOT of \sysname is similar to that of the temporal-spatial sharing method, MuxServe, and the dedicated method, vLLM.
\revise{
Note that the Avg TTFT of MuxServe is the worst mainly due to two reasons.
First, its RR scheduling policy for the prefill phase increases the TTFT.
Second, MuxServe restricts the used percentage of GPU computation units, which can greatly prolong the compute-bound prefill phase.
}

\begin{figure}[t]
	\begin{minipage}{0.49\linewidth}
        \vspace{-0.1in}
        \centerline{\includegraphics[width=\linewidth,trim=0 0 0 10, clip]{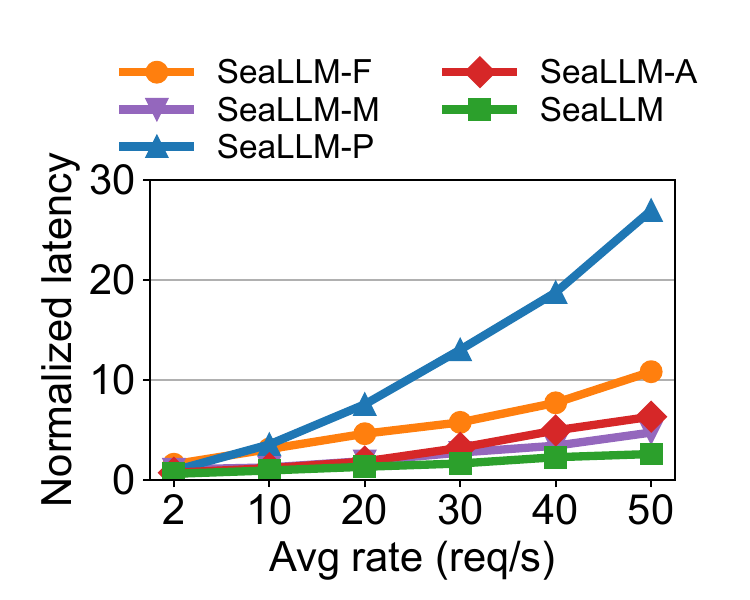}}
        \vspace{-0.05in}
        \centerline{\small (a) Normalized latency.}
        \vspace{-0.05in}
    \end{minipage}
	\begin{minipage}{0.49\linewidth}
        \vspace{-0.1in}
        \centerline{\includegraphics[width=\linewidth,trim=0 0 0 10, clip]{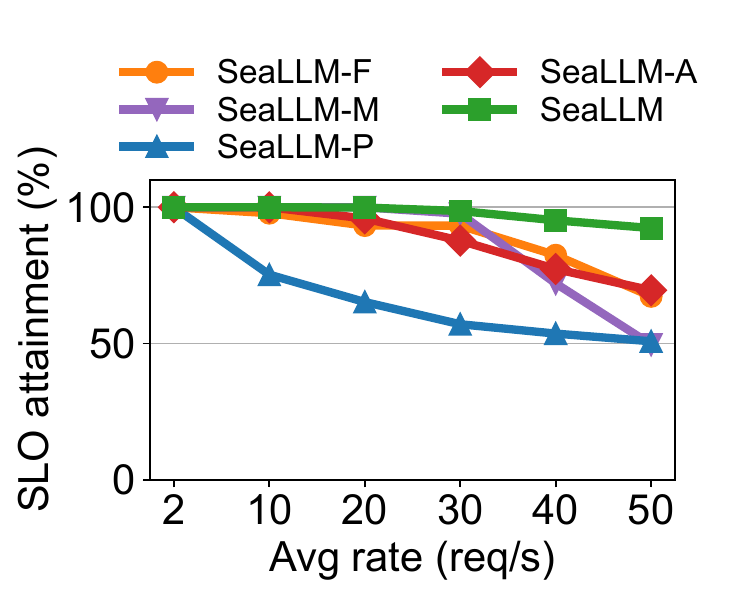}}
        \vspace{-0.05in}
        \centerline{\small (b) SLO attainment.}
        \vspace{-0.05in}
    \end{minipage}
    \caption{\revise{Impact of the \sysname design choices.}}
    \label{fig:eval_design}
    \vspace{-0.1in}
\end{figure}

\subsection{Ablation Study}
\paraf{Impact of \sysname design.}
To show the efficacy of our design choices, we evaluate \sysname with four variants:
(1) \sysname using the FCFS scheduling algorithm (\sysname-F);
\revise{
(2) \sysname using the skip-join MLFQ scheduling algorithm used in FastServe~\cite{wu2023fast} (\sysname-M);
}
(3) \sysname without the placement algorithm (\sysname-P);
and (4) \sysname without the adaptive replacement algorithm (\sysname-A);
We use the MAF traces and vary the average request rate from 2 to 50 req/s.
\revise{
\sysname-F uses the widely used FCFS scheduling algorithm, while \sysname-M uses one of the SOTA scheduling algorithms for single LLM services.
These two variants have up to $4.17\times$ larger normalized latency and $1.37\times$ lower SLO attainment.
Our latency-optimized scheduling algorithm outperforms FCFS because \sysname alleviates the head-of-line blocking problem, and outperforms skip-join MLFQ because \sysname integrates the service characteristics into the scheduling policy.
}
\sysname-P does not share LLM services and thus is degraded to vLLM.
\sysname-A uses the fixed placement and cannot change the placement according to the request traffic.
\sysname-A increases the normalized latency by up to $2.44\times$ and decreases the SLO attainment by up to $1.33\times$, showing the benefit of our adaptive replacement algorithm.

\begin{figure}[t]
    \begin{minipage}[t]{0.42\linewidth}
    \centerline{\includegraphics[width=0.9\linewidth]{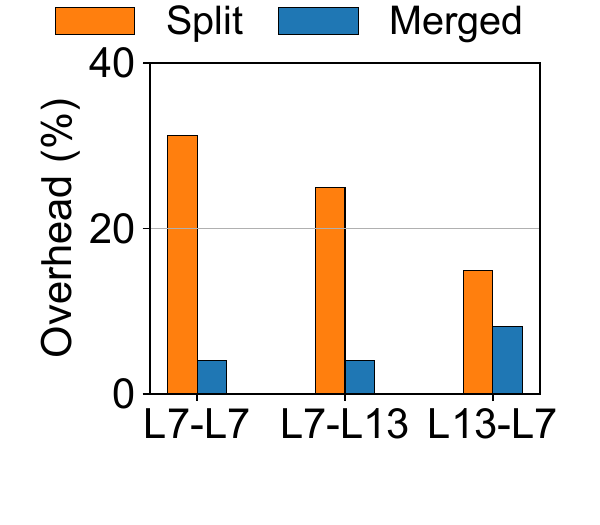}}
    \vspace{-0.2in}
    \caption{Impact of the unified KV cache.}
    \vspace{-0.1in}
    \label{fig:eval_kvcache}
    \end{minipage}
    \begin{minipage}[t]{0.51\linewidth}
    \centerline{\includegraphics[width=0.9\linewidth]{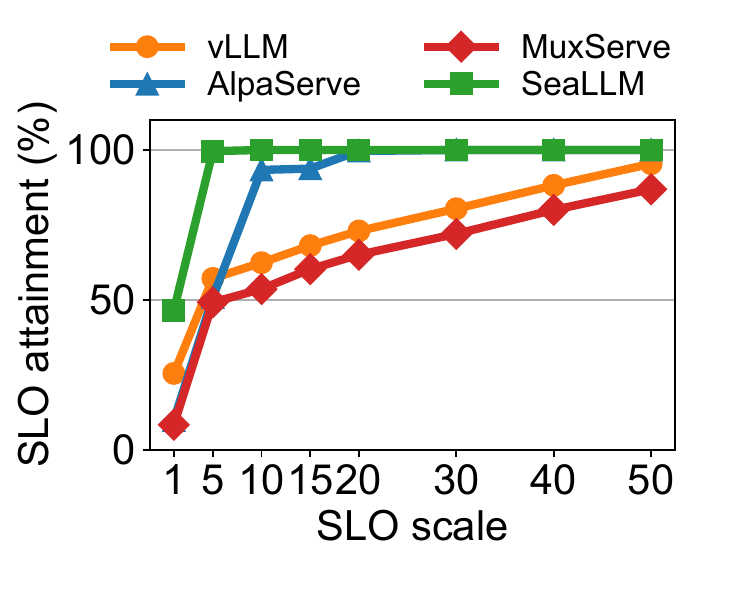}}
    \vspace{-0.2in}
    \caption{\revise{Impact of the SLO scale.}}
    \vspace{-0.1in}
    \label{fig:eval_slo}
    \end{minipage}
\end{figure}

\begin{figure}[t]
	\begin{minipage}{0.49\linewidth}
        \centerline{\includegraphics[width=\linewidth,trim=0 0 0 30, clip]{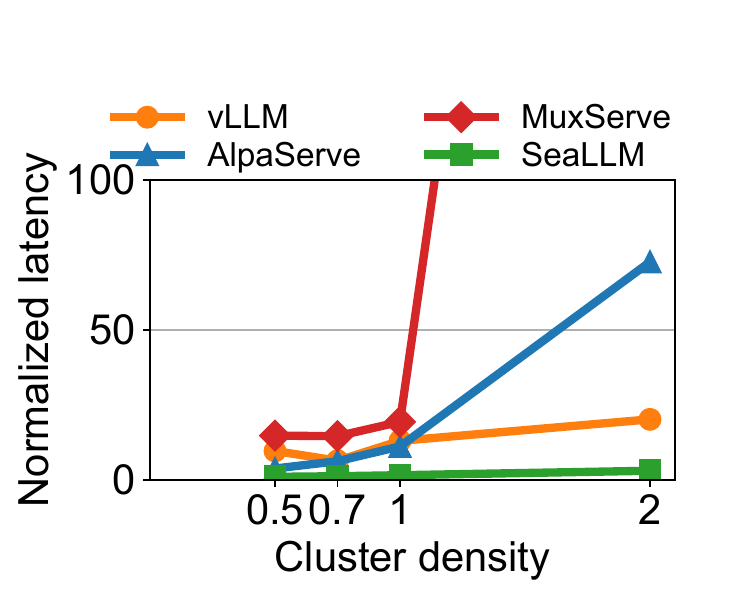}}
        \vspace{-0.05in}
        \centerline{\small (a) Normalized latency.}
        \vspace{-0.05in}
    \end{minipage}
	\begin{minipage}{0.49\linewidth}
        \centerline{\includegraphics[width=\linewidth,trim=0 0 0 30, clip]{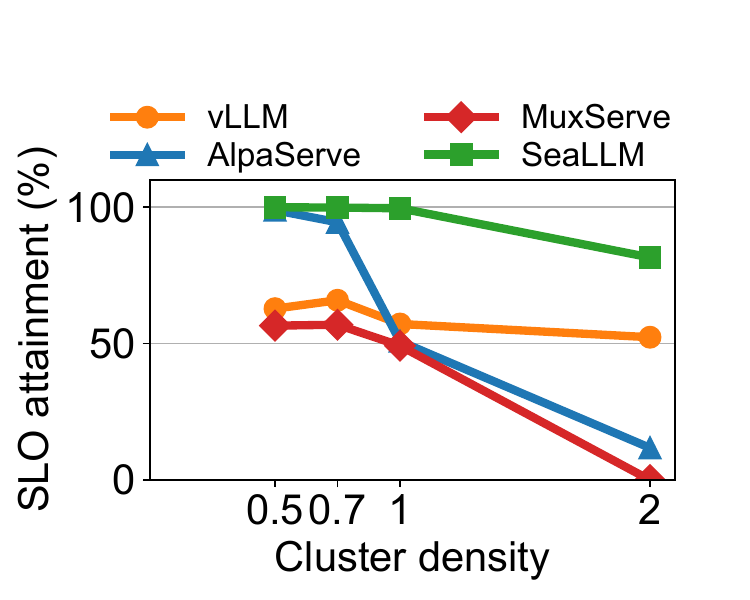}}
        \vspace{-0.05in}
        \centerline{\small (b) SLO attainment.}
        \vspace{-0.05in}
    \end{minipage}
    \caption{Impact of the cluster density.}
    \label{fig:eval_density}
    \vspace{-0.1in}
\end{figure}

\revise{
To study the effectiveness of \sysname's unified KV cache, we compare the overhead of the multi-LLM executors with the merged cache block and the split cache block as shown in Figure~\ref{fig:eval_kvcache}.
A-B represents the overhead of running LLM A when sharing it with LLM B.
L7 and L13 are short for Llama2-7B and Llama2-13B, respectively.
We select three cases in which the used LLMs in L7-L7 have the same block size and the used LLMs in L7-L13 and L13-L7 have different block sizes.
\sysname's merged cache block consistently brings lower overhead than the split cache block.
The reason is that the merged cache block uses smaller block table and has better read/write locality than the split one.
}

\parabf{Impact of the SLO scale.}
\revise{
We also evaluate \sysname under different SLO scales as shown in Figure~\ref{fig:eval_slo}.
The SLO for each LLM service is set to SLO scale times the average execution time of each LLM service.
\sysname always reaches better SLO attainment than other baselines.
The improvement shows the generality of \sysname with different SLO strictness.
Besides, the lower SLO attainments of baselines indicate that they need clusters with more GPUs than \sysname to meet the SLO requirements.
In other words, \sysname has better resource utilization and serving performance than baselines.
}

\parabf{Impact of the cluster density.}
To investigate the impact of the cluster density, we evaluate \sysname over different cluster sizes with the same LLM services and request trace as shown in Figure~\ref{fig:eval_density}.
Densities $0.5-2$ represent clusters with 64 GPUs, 48 GPUs, 32 GPUs, and 16 GPUs, respectively, where a larger density number represents a more crowded cluster.
Overall, \sysname has better performance than all the baselines, especially on the more crowded cluster.
The performance of AlpaServe degrades fast with larger cluster density because its placement algorithm cannot find the optimal placement on crowded clusters.
MuxServe cannot find available placement plans when the cluster density is 2.

\begin{figure}[t]
	\begin{minipage}{0.49\linewidth}
        \centerline{\includegraphics[width=\linewidth,trim=0 0 0 30, clip]{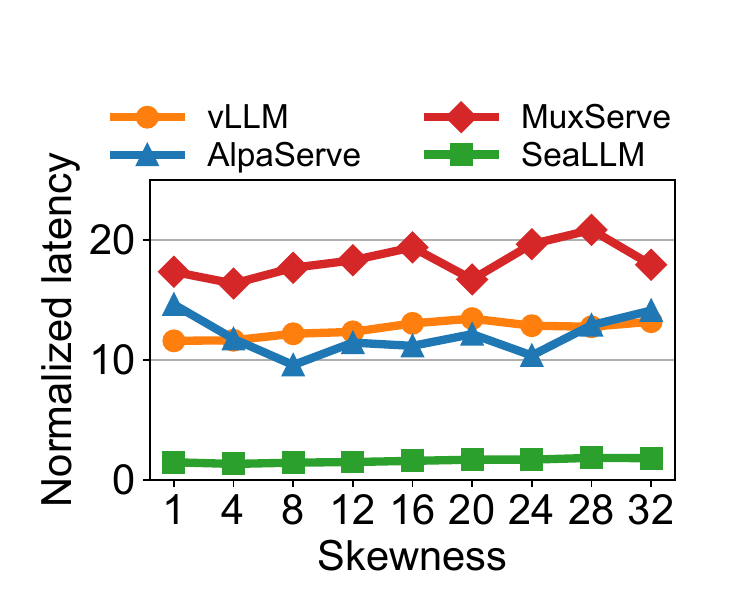}}
        \vspace{-0.05in}
        \centerline{\small (a) Normalized latency.}
        \vspace{-0.05in}
    \end{minipage}
	\begin{minipage}{0.49\linewidth}
        \centerline{\includegraphics[width=\linewidth,trim=0 0 0 30, clip]{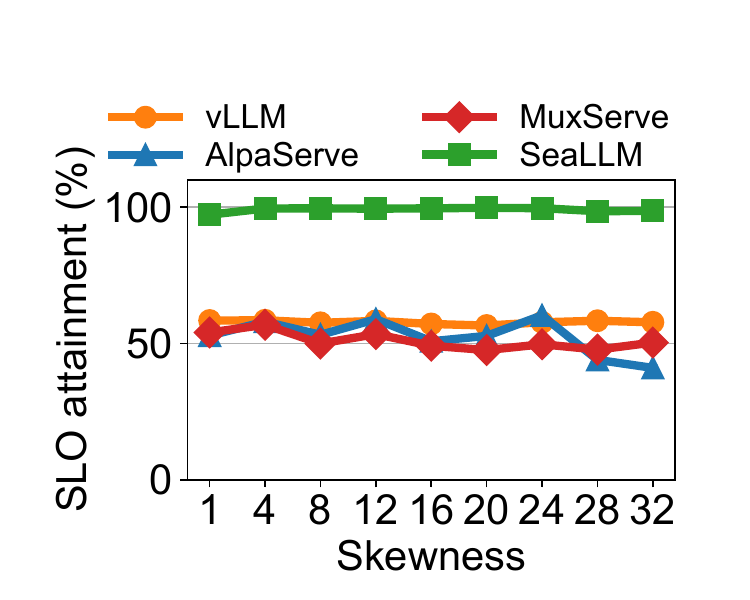}}
        \vspace{-0.05in}
        \centerline{\small (b) SLO attainment.}
        \vspace{-0.05in}
    \end{minipage}
    \caption{Impact of the workload skewness.}
    \label{fig:eval_skewness}
    \vspace{-0.1in}
\end{figure}

\begin{figure}[t]
	\begin{minipage}{0.49\linewidth}
        \vspace{-0.05in}
        \centerline{\includegraphics[width=\linewidth,trim=0 0 0 30, clip]{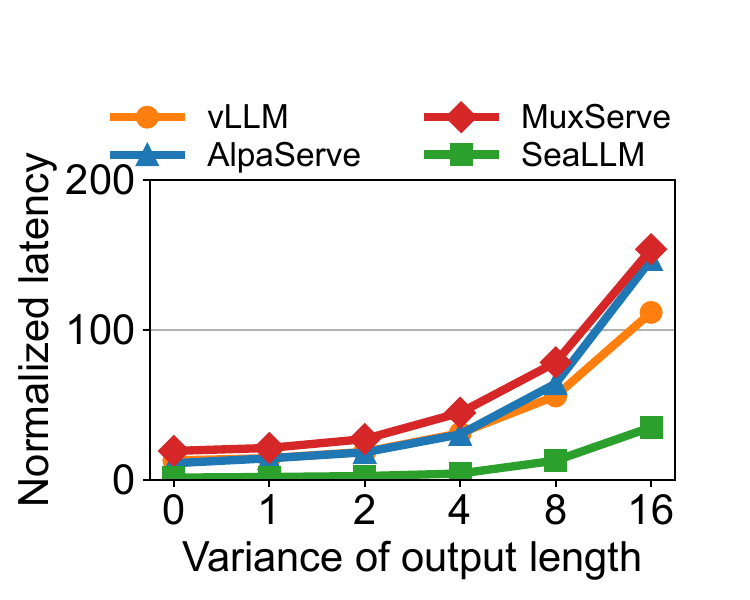}}
        \vspace{-0.05in}
        \centerline{\small (a) Normalized latency.}
        \vspace{-0.05in}
    \end{minipage}
	\begin{minipage}{0.49\linewidth}
        \vspace{-0.05in}
        \centerline{\includegraphics[width=\linewidth,trim=0 0 0 30, clip]{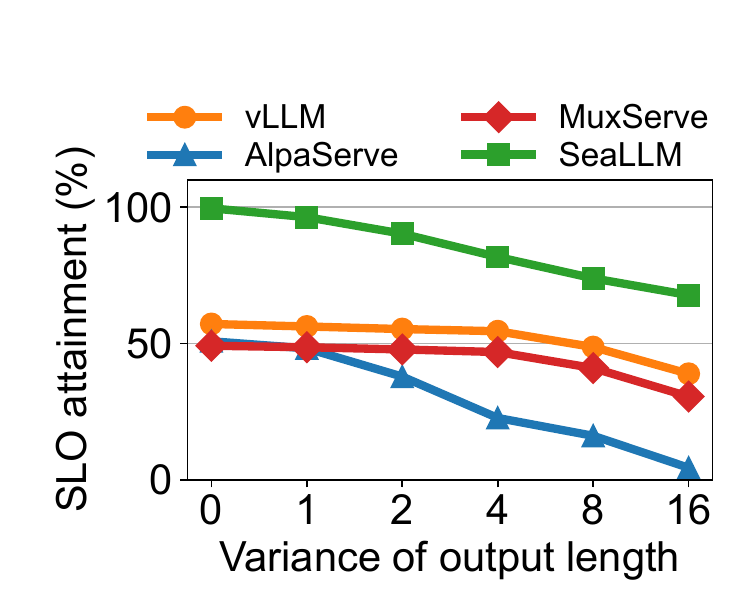}}
        \vspace{-0.05in}
        \centerline{\small (b) SLO attainment.}
        \vspace{-0.05in}
    \end{minipage}
    \caption{Impact of the profiling accuracy.}
    \vspace{-0.05in}
    \label{fig:eval_profile_acc}
\end{figure}

\parabf{Impact of the workload skewness.}
We evaluate \sysname with different workload skewness as shown in Figure~\ref{fig:eval_skewness}.
We use the number of consecutive requests from the same service as the skewness.
More consecutive requests from the same service represent larger workload skewness.
\sysname has the best performance among all baselines under different skewness.
Besides, the performance for all baselines is slightly decreased when the skewness increases.
This is because larger skewness means more bursty requests, which is more challenging for the serving system.
The metric curves of \sysname do not fluctuate much, illustrating that \sysname is good at serving skew LLM services and bursty requests.

\parabf{Impact of the profiling accuracy.}
As \sysname relies on the profiled service characteristics to make scheduling decisions, we also study the impact of the profiling accuracy.
For each request, we sample the delta length from a Gaussian distribution with average 0 and standard variance from 0 to $16\times$ of the output length.
Figure~\ref{fig:eval_profile_acc} shows that all methods perform worse under a high variance of the output length, because there are longer requests under high variance and they can impair the performance greatly.
The benefits of \sysname get smaller under a high variance of the output length because the inaccurate profiling information affects our scheduling algorithm.
However, \sysname still outperforms other baselines under different variances of the output length, showing the efficacy and robustness of our system design.

\subsection{System Overhead}

\begin{figure}[t]
    \begin{minipage}[t]{0.48\linewidth}
    \centerline{\includegraphics[width=0.9\linewidth]{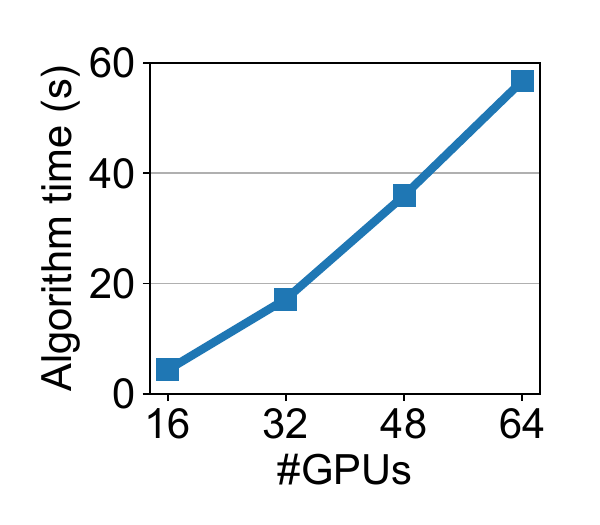}}
    \vspace{-0.25in}
    \caption{\revise{The running time of our placement algorithm.}}
    \vspace{-0.1in}
    \label{fig:eval_overhead}
    \end{minipage}
    \begin{minipage}[t]{0.48\linewidth}
    \centerline{\includegraphics[width=0.9\linewidth]{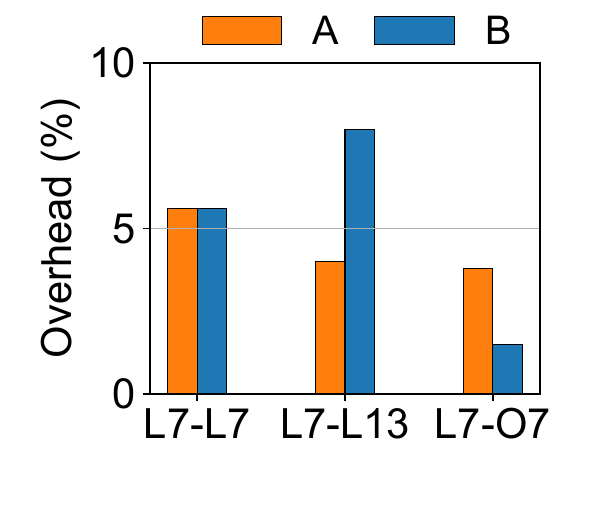}}
    \vspace{-0.25in}
    \caption{The overhead of the multi-LLM executor.}
    \vspace{-0.1in}
    \label{fig:eval_executor}
    \end{minipage}
\end{figure}

\paraf{Placement algorithm running time.}
We evaluate the running time of our placement algorithm with different GPU numbers, as shown in Figure~\ref{fig:eval_overhead}.
We can speed up the placement algorithm with at most eight parallel processes, considering the possible TP sizes and the number of shared LLM services.
The running time is shorter than one minute, which can be fully overlapped by running before the replacement moment.

\parabf{Multi-LLM executor.}
Figure~\ref{fig:eval_executor} shows the increase in iteration time when sharing multiple LLMs on the same set of GPUs.
A-B represents sharing LLM service A with LLM service B.
L7, L13, and O7 are short for Llama2-7B, Llama2-13B, and OPT-6.7B, respectively.
We select three cases to demonstrate that the overhead is acceptable.
The three cases are sharing LLMs with the same architecture (L7-L7), sharing LLMs with similar architecture but different model sizes (L7-L13), and sharing LLMs with different architectures (L7-O7).
The increases in iteration time are under 4\% for three out of six LLMs and under 8\% for all cases.

%% file: sections/related.tex
\section{Related Work}
\label{sec:related}

\paraf{LLM serving systems.}
The fast development of LLMs has drawn great attention to LLM serving systems.
Some systems leverage efficient operators to improve the execution speed of LLMs, 
e.g., TensorRT-LLM~\cite{NVIDIA2023trtllm}, LightSeq~\cite{wang2020lightseq}, Flash Attention~\cite{dao2022flashattention}, and Flash-Decoding~\cite{hong2024flashdecoding}.
Recent work, like FastTransformer~\cite{NVIDIA2019FTT}, Deepspeed-Inference~\cite{aminabadi2022deepspeed}, and TGI~\cite{HuggingFace2023tgi}, has explored intra- and inter-operator parallelism for fast serving.
Orca~\cite{yu2022orca} and FastServe~\cite{wu2023fast} propose scheduling algorithms for LLMs.
Besides, SplitWise~\cite{patel2024splitwise}, DistServe~\cite{zhong2024distserve}, and TetriInfer~\cite{hu2024inference} propose disaggregation of the prefill phase and decoding phase of LLM serving.
vLLM~\cite{kwon2023efficient} manages the memory usage of LLMs.
LoongServe~\cite{wu2024loongserve}, SARATHI~\cite{agrawal2023sarathi}, and DeepSpeed-FastGen~\cite{holmes2024deepspeed} are optimized for long-context settings.
These systems optimize the execution of a single LLM.
\revise{
Llumnix~\cite{biao2024llumnix} notices the dynamic request traffic and reschedules requests among multiple model instances.
However, Llumnix optimizes multiple instances of the same LLM service.
\sysname explores the multiple-LLM-service setting which is orthogonal to these single-LLM optimizations.
}

\parabf{GPU sharing.}
GPU sharing has been widely studied for DL models.
Some existing approaches~\cite{zhao2022multi,xiao2020antman,lim2021zico} 
are designed for training DL models, which usually have a repeated pattern.
They are not suitable for sharing serving workloads which are time-varying and bursty.
Some systems~\cite{han2022microsecond,li2023alpaserve} are designed for sharing serving workloads.
These approaches do not consider the autoregressive pattern of LLM serving and directly applying them to LLM services can cause the head-of-line blocking problem.
dLoRA~\cite{wu2024dlora} is designed for sharing LLM services.
However, it is specifically designed for the multi-LoRA scenario which cannot be applied to general scenarios.
MuxServe~\cite{duan2024muxserve} is the closest work to \sysname, but it utilizes a scheduling algorithm combining RR and FCFS, which increases the request latency and the memory burden of GPUs.
Besides, MuxServe optimizes the system throughput and preallocates GPU quota for each LLM service, which cannot react timely to the fluctuations in request traffic.
\sysname is designed for sharing general LLM services.
Besides, \sysname utilizes a latency-optimized scheduling algorithm to minimize normalized latency and an adaptive replacement algorithm for the changing request traffic.

\parabf{Model parallelism.}
As the model size of LLMs is extremely large, model parallelism is required for training and serving LLMs.
Tensor parallelism~\cite{shoeybi2019megatron,rajbhandari2020zero} splits operators on different devices and is widely used for LLM workloads.
Besides, pipeline parallelism~\cite{huang2019gpipe, li2021terapipe, narayanan2019pipedream} splits LLM layers on different devices. 
As the context is getting longer, sequence parallelism is introduced to LLM 
workloads~\cite{li2021sequence, korthikanti2023reducing,wu2024loongserve}.
The model parallelisms are usually mixed for LLM training, while for LLM serving, the most widely used parallelism pattern is tensor parallelism.
Note that the placement algorithm of \sysname is not limited to tensor parallelism,
and \sysname can integrate new parallelism techniques into its placement algorithm.

%% file: sections/conclusion.tex
\section{Conclusion}
\label{sec:conclusion}

In this paper, we presented \sysname, a service-aware and latency-optimized sharing system for multiple LLMs.
\sysname uses a latency-optimized scheduling algorithm by utilizing the service characteristics.
\sysname exploits a placement algorithm to find the placement plan and utilizes an adaptive replacement algorithm to determine the replacement interval.
To execute shared LLMs, we propose the unified KV cache to manage the memory resources.
Evaluation with real-world traces and LLM services shows that \sysname significantly improves the normalized latency, tail latency, and SLO attainment compared to prior SOTA solutions.

%% file: sections/appendix.tex
\appendix
\section{Appendix}
\subsection{Placement Algorithm}
\label{sec:appendix_placement}
Algorithm~\ref{alg:placement} shows the pseudocode of \sysname's placement algorithm.

\begin{algorithm}[h]
  \caption{Placement algorithm}
  \begin{algorithmic}[1]
  \begin{small}
  \Statex \textbf{Input:} LLM services $S$; cluster GPUs $G$; requests $R$.
  \begin{algorithmic}[1]
      \State $P^*\gets \emptyset$
      \State $\{p'\}\gets GetParallelismConfig(G,S)$
      \State \textbf{// Enumerate GPU groups}
      \For{$p\in \{p'\}$}
          \State $P\gets \emptyset$
          \State $B\gets GenerateGPUGroups(G, p)$
          \State \textbf{// Allocate LLM services}
          \While{True}
              \State $UI\gets SimulateUI(R, P)$
              \State $FindFlag\gets False$
              \For{$b\in SortedByAvgRate(B)$}
                  \For{$s\in SortedByUI(UI, S)$}
                      \If{$b.CanAllocate(s)$}
                          \State $P.AddServices(b, s)$
                          \State $FindFlag\gets True$
                          \State $Break$ \textbf{// To Line 17}
                      \EndIf
                  \EndFor
              \EndFor
              \If{$FindFlag==False$}
                  \State $Break$
              \EndIf
          \EndWhile
          \State \textbf{// Simulate and get the optimal placement}
          \State $M_{est}\gets SimulatePerformance(R, P)$
          \If{$M_{est}.SLO>P^*.SLO$}
              \State $P^*\gets P$
          \ElsIf{$M_{est}.SLO=P^*.SLO$ and \\ \ \ \ \ \ \ \ \ \ \ \ \ \ \ \ \ \  $M_{est}.norm\_latency<P^*.norm\_latency$}
              \State $P^*\gets P$
          \EndIf
      \EndFor
  \end{algorithmic}
  
  \end{small}
  \end{algorithmic}
  \label{alg:placement}
  \end{algorithm}

\subsection{Scheduling Algorithm}
\label{sec:appendix_schedule}
\begin{algorithm}[h]
  \caption{DB scheduling algorithm}
  \begin{algorithmic}[1]
  \begin{small}
  
  \Statex \textbf{Input:} Requests $R$; LLM services $S$.
  
  \begin{algorithmic}[1]
      \State \textbf{// Initialize}
      \State $wait\_queue\gets PriorityQueue$
      \State $decoding\_queue \gets PriorityQueue$
      \While{$True$}
          \State \textbf{// Process newly arrived requests}
          \While{a new request $r_{new}$ arrives}
              \State $InitBudget(r_{new})$
              \State $InitPriority(r_{new})$
              \State $wait\_queue.AddRequest(r_{new})$
          \EndWhile
          \State \textbf{// Schedule requests}
          \State $to\_run\_batch\gets \emptyset$
          \State $r\gets GetHighestPriorityOrStarvedRequest()$
          \If{$r\in wait\_queue$}
              \State $cur\_queue \gets wait\_queue$
          \Else
              \State $cur\_queue \gets decoding\_queue$
          \EndIf
          \While{$EnoughResource()$}
              \If{$r.service=to\_run\_batch.service$}
                  \State $to\_run\_batch.AddRequest(r)$
                  \State $r\gets cur\_queue.Get()$
              \EndIf
          \EndWhile
          \State $Execute(to\_run\_batch)$
          \State $UpdateBudget(to\_run\_batch)$
          \State $UpdatePriority(to\_run\_batch)$
      \EndWhile
  \end{algorithmic}
  
  \end{small}
  \end{algorithmic}
  \label{alg:db}
  \end{algorithm}

Algorithm~\ref{alg:db} shows the pseudocode of \sysname's DB scheduling algorithm.
The proof of Theorem~\ref{thm:schedule} is as follows,

\begin{proof}
Our placement algorithm shares at most two services for each sharing group, and assume they are $s_0$ and $s_1$.
The execution time on one GPU for $s_0$ is $\hat{L_0}$ and for $s_1$ is $\hat{L_1}$.
Assume one execution order $R'$ is better than the optimal execution order $R^*$, and the scheduling decisions are different for only two requests $j$ and $k$ at time $t$.
Specifically, for time $t$, request $j$ is decided to execute in $R^*$, but not in $R'$.
Request $k$ is decided to execute in $R'$, but not in $R^*$.
Assume the left time for $j$ is $T_j$ and for $k$ is $T_k$.
$j$ is submitted at $t_j$ and $k$ is submitted at $t_k$.
According to the sorting rule of $R^*$, we can get,
\begin{equation}
\label{equ:proof_base}
T_j \hat{L_j} < T_k \hat{L_k}.
\end{equation}
The difference between the normalized latencies of $R'$ and $R^*$ is,
\begin{equation}
\label{equ:proof_diff}
\begin{aligned}
D &= (\frac{t+T_k-t_k}{\hat{L_k}}+\frac{t+T_k+T_j-t_j}{\hat{L_j}}) \\ 
  &\ \ \ \   - (\frac{t+T_j-t_j}{\hat{L_j}} + \frac{t+T_j+T_k-t_k}{\hat{L_k}}) \\
  &= \frac{T_k}{\hat{L_j}}-\frac{T_j}{\hat{L_k}} \\
  &< 0,
\end{aligned}
\end{equation}
where we only consider requests $j$ and $k$ because other requests have the same execution decisions.
From Equation~\ref{equ:proof_diff}, we can get $T_j \hat{L_j} > T_k \hat{L_k}$, which is contradicted to Equation~\ref{equ:proof_base}.
If there are more than two requests having different execution orders, we can prove them pair by pair.
To conclude, $R^*$ is the optimal execution order.
\end{proof}

The proof of Theorem~\ref{thm:schedule_prio} is as follows,

\begin{proof}
\begin{lemma}
\label{lemma:cmu_rule}
For a parallel single server queueing system with infinite buffer, the optimal policy is the priority policy based on the c-$\mu$ rule: The server selects the request $r$ in queue $r^*=argmax\{c_r \mu_r \}$, where $c_r$ is the marginal delay cost and $1/\mu_r$ is the average processing time of request $r$.
\end{lemma}

Lemma~\ref{lemma:cmu_rule} is proven in ~\cite{van1995dynamic,tran2022finding}.
In our scenario, the delay cost function of request $i$ that resides $\tau$ units of time is 
\begin{equation}
C_i(\tau)=\tau/\hat{L}_r.
\end{equation}
The marginal delay cost function can be calculated as 
\begin{equation}
c_i = C'_i(\tau) = 1/\hat{L}_r.
\end{equation}
For any scheduling moment, the average processing time of request $i$ is the estimated remaining time $T'_r$.
Consequently, the optimal priority is
\begin{equation}
\begin{aligned}
c_i*\mu_r &= 1/\hat{L}_r * 1/T'_r \\
                    &= 1/(T'_r \hat{L}_r).
\end{aligned}
\end{equation}
From the above equation, we should serve the request with the maximum $1/(T'_r \hat{L}_r)$ first, i.e., minimum $T'_r \hat{L}_r$.
To conclude, our scheduling priority $O_r = T'_r \hat{L}_r$ is the optimal priority to minimize the expected normalized latency $\mathbb{E}(L_N)$ for any scheduling moment.
\end{proof}